\documentclass[a4paper,fleqn,usenatbib]{mnras}

\usepackage{newtxtext,newtxmath}

\usepackage[T1]{fontenc}
\usepackage{ae,aecompl}

\usepackage{graphicx}	
\usepackage{amsmath}	
\usepackage{amssymb}	

\title[]{A new look at Sco OB1 association with Gaia DR2}

\author[Lidia Yalyalieva et al.]{
	L. Yalyalieva,$^{1,2,3}$\thanks{E-mail: yalyalieva@yandex.ru}\
	G. Carraro,$^{3}$\
	R. Vazquez,$^{4}$\
	L. Rizzo,$^{4}$\	
	E. Glushkova,$^{1,2}$\
	and E. Costa$^{5}$\\
	$^{1}$Physics Department, Lomonosov Moscow State University, Leninskie Gory, Moscow, 119991, Russian Federation\\
	$^{2}$Sternberg Astronomical Institute, Lomonosov Moscow State University, Universitetsky pr.13, Moscow, 119234, Russian Federation\\
	$^{3}$Department of Physics and Astronomy, Padova University, Vicolo Osservatorio 3, I-35122, Padova, Italy\\
	$^{4}$Instituto de Astrof\'{\i}sica de La Plata (CONICET, UNLP),
              Paseo del Bosque s/n, La Plata, Argentina \\
         $^{5}$Departamento de Astronomia, Universidad de Chile, Casilla 36-D, Santiago, Chile
              }

\date{Accepted 2020 April 25. Received 2020 April 23; in original form 2020 February 19
}

\pubyear{2020}

\begin{document}
	\label{firstpage}
	\pagerange{\pageref{firstpage}--\pageref{lastpage}}
	\maketitle
	
	\begin{abstract}
	We present and discuss photometric optical data in the area of the OB association Sco OB1\ covering about 1 squared degree. UBVI photometry is employed in tandem with Gaia DR2 data to investigate the 3 dimensional structure and the star formation history of the region. By combining parallaxes and proper motions we identify 7 physical groups located between the young open cluster NGC 6231 and the bright nebula IC4628. The most prominent group coincides with the sparse open cluster Trumpler 24. We confirm the presence of the intermediate age star cluster VdB-Hagen 202, which is unexpected in this environment, and provide for the first time estimates of its fundamental parameters. After assessing individual groups membership, we derive mean proper motion components, distances, and ages. The seven groups belong to two different families. To the younger family (family I) belong several pre-Main Sequence stars as well. These are evenly spread across the field, and also in front of VdB-Hagen 202. VdB-Hagen 202 and two smaller, slightly detached, groups of similar properties form family II, which do not belong to the association, but are caught in the act of passing through it. As for the younger population, this forms an arc-like structure from the bright nebula IC 4628 down to NGC 6231, as previously found. Moreover, the pre-Main Sequence stars density seems to increase from NGC 6231 northward to Trumpler 24. \end{abstract}
	
	
	
	
	\section{Introduction}
	
	According to \citep{Lada03} most stars form in relatively compact clusters with more than 100 members.  In this scenario stellar associations,
	namely loose groups of stars of early spectral types, are interpreted as  the early stages of dynamical dissolution of star clusters. This process is referred to as {\textit infant mortality}, and it is triggered by gas expulsion from stellar winds and, probably, SNae events. Therefore, stars clusters would form embedded in molecular clouds and bound, but the largest part of them will quickly expand and then slowly dissolve.
	This paradigm has been recently challenged by a detailed  study of nearby stellar associations using Gaia DR2 data \citep{Ward2019}. Most associations in fact do not show any expansion signatures confirming earlier results by \citep{Mel2018}.
	Besides, associations are unbound loose ensembles containing a sizeable fraction of OB stars \citep{efre1989}, and are organised in fractal structures
	\citep{goul2018}, going from associations to aggregate, and then complexes and super-complexes.
	Their existence, young age,  and structure should contain imprints of their recent formation.
	Therefore, detailed studies of individual star 3-dimensional
	structure,  age, and kinematics are essential to understand their formation and, in turn, to constrain the star formation process better \citep{Beccari}.\\
	
\begin{figure}
		\includegraphics[width=\columnwidth]{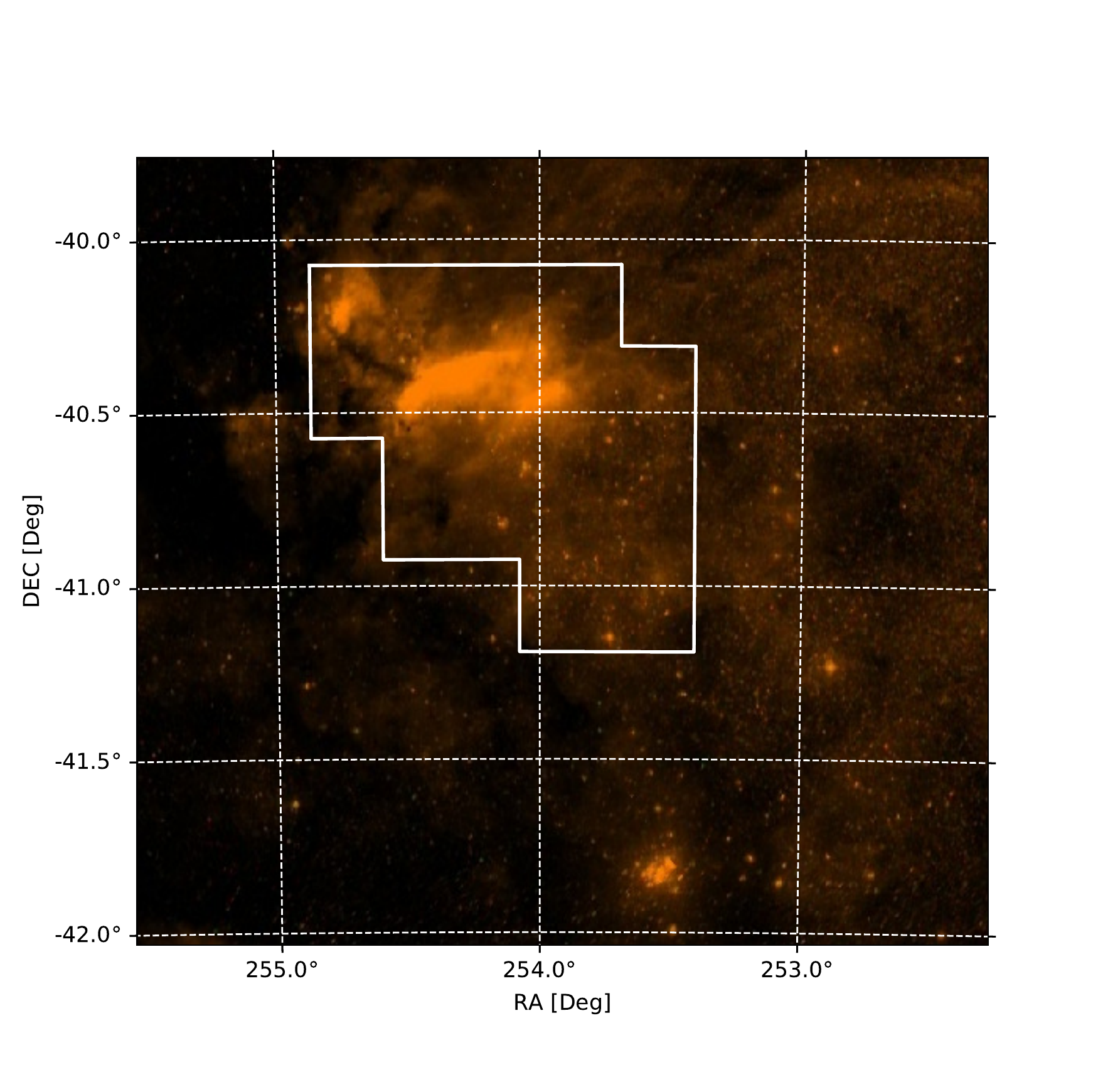}
		\caption{A Digital Sky Survey image of the Sco OB1 area. The white solid polygon encloses the field covered by our photometry (see also Fig. \ref{fig:Fig2}). The conspicuous star cluster in the south-west is NGC 6231, while the norther part is dominated by the \ion{H}{ii} region G345.45+1.50. Several bright, probably early type stars are spread across the field. }
		\label{fig:Fig1}
	\end{figure}

	\noindent
	With the aim of providing additional  insights on this important topic,  in this paper  we investigate an  area of one degree on a side in the Sco OB1 association, centred at  $\alpha = 253\degr.97$, $\delta = -40\degr.64$ (Fig.~\ref{fig:Fig1}), for which we secured multi-band optical photometry.
Sco1 is a very rich and complex stellar association \citep{Damiani}.
The spectacular  \ion{H}{ii} region G345.45+1.50  is situated in the northern part of the field, while the most prominent young star cluster, NGC 6231 \citep{Baume,Feinstein,Sung98,Sung}, is located in the southern part. We are not covering this cluster in our study, but we are concentrating on the northern and central region.
Here, the most interesting structure  is Trumpler~24. This is thought to be a young  open cluster with poorly defined boundaries and  complex structure belonging to Sco OB1 \citep{Heske}. 
Besides, the area under investigation is rich in pre-main sequence stars \citep{Heske, Damiani}, which indicates active/recent star formation. \\

\noindent
Because of the presence of a lot of gas irregularly distributed which causes a significant differential reddening, 
photometry is not  enough to identify  known and/or unknown stellar groups and determine their 
population.  For this reason, we complement our photometric data-set with  the high quality astrometric data from Gaia DR2 \citep{Gaia} .        
To this aim we recall the reader that
recently a number of special tools and approaches for group searching and assigning membership probabilities was constructed, including clustering methods \citep{Krone-Martins} and unsupervised learning algorithms \citep{Tang,Yuan}. Initial data could be of different types: researches could use only photometry \citep{Buckner}, only astrometry \citep{Cantat-Gaudin} or combine these two sources \citep{Krone-Martins}.\\
	
\noindent
Therefore, the layout of the paper is as follow:  in Section~2 we describe our photometric observation. Section~3 is dedicated to the detection of stellar groups in the area. We discuss the population of PMS stars in the region in Section~4 Finally, Section~5 summarises our findings.

	\begin{figure}
		\includegraphics[width=\columnwidth]{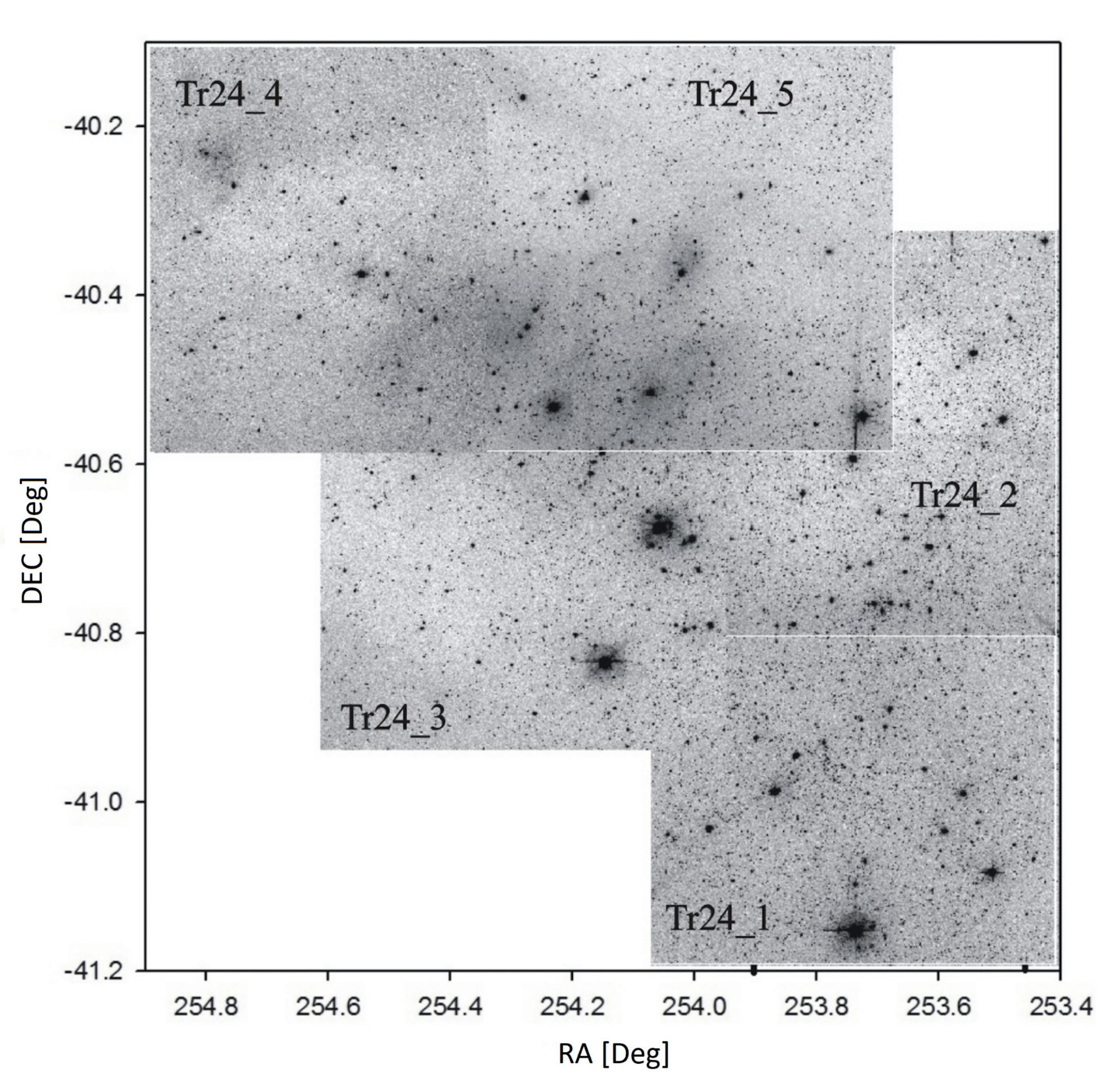}
		\caption{Calibrated CCD frames, numbered as in Table \ref{tab:t1}, to show the quality of the images and the actual field coverage. }
		\label{fig:Fig2}
	\end{figure}
		
\section{Observations and data reduction} 
 
Trumpler 24 was originally observed at Las Campanas Observatory (LCO) on the nights of August 12 and 13, 2016, using the 1-m Henrietta Swope Telescope\footnote{http://www.lco.cl/telescopes-information/herrietta-swope}. The camera employed for direct CCD imaging in the $UBVI$ pass-bands was the E2V CCD231-\#84 one ($4096\times4112$ pixels), with a scale of 0.435$^{\prime\prime}$/pixel and a field-of-view (FOV) of $29,7^{\prime}\times29,8^{\prime}$. The CCD was operated without binning at a nominal gain of 1.04 $e^{-}$/ADUs, implying thus a readout noise of 3.4$e^{-}$ per quadrant.\\
Along the two observing runs, named N1 and N2, we needed five target fields to properly cover the whole area of Trumpler 24; they are indicated Tr24\_1, Tr24\_2, etc, in Fig.~\ref{fig:Fig2} and Table~\ref{tab:t1}, respectively. As can be seen in Fig.\ref{fig:Fig2} there is enough overlapping ($3^{\prime}$ to $5^{\prime}$) amongst the different frames. The observations were carried out under reasonable seeing conditions (always less than $1.5^{\prime\prime}$).\\

The transformation from instrumental system to the standard Johnson-Kron-Cousins system, as well as corrections for atmospheric extinction, were determined through multiple observations of standard stars in Landolt areas G93-48, PG2213-006, and TPHE \citep{lan2009}. To this aim,  these stars were observed at air mass ranging from 1.06 to 2.12 approximately for both nights. 
As for the colour coverage, the standard stars sweep a range of -0.29 $ \leq B-V \leq$ 1.55 and -1.22$ \leq U-B \leq$ 1.87, therefore very suitable for a young stellar region as Trumpler 24 is assumed to be.\\

\begin{table}
\caption{\label{tab:t1}$UBVI$ photometric observations of the 5 target fields.}
\centering
\begin{tabular}{lcccc}
\hline\hline
Target&Date&Filter&Exposure(s)&airmass\\
\hline

Tr24\_1&N1&$U$&60,300&1.035~-~1.038\\
&&$B$&30,200&1.042~-~1.043\\
&&$V$&15,150&1.046~-~1.048\\
&&$I$&15,100&1.029~-~1.030\\
\hline
Tr24\_2&N2&$U$&60,300&1.020~-~1.021\\
&&$B$&60,200&1.020~-~1.021\\
&&$V$&15,100&1.020~-~1.020\\
&&$I$&15,100&1.021~-~1.021\\
\hline
Tr24\_3&N1&$U$&60,300&1.024~-~1.025\\
&&$B$&30,200&1.021~-~1.021\\
&&$V$&15,100&1.022~-~1.022\\
&&$I$&15,100&1.026~-~1.026\\
\hline
Tr24\_4&N2&$U$&60,300&1.024~-~1.024\\
&&$B$&60,200&1.021~-~1.021\\
&&$V$&15,100&1.022~-~1.022\\
&&$I$&15,100&1.026~-~1.026\\
\hline
Tr24\_5&N1&$U$&60,300&1.020~-~1.021\\
&&$B$&30,200&1.019~-~1.019\\
&&$V$&15,100&1.020~-~1.021\\
&&$I$&15,100&1.022~-~1.023\\

\hline
\end{tabular}
\end{table}

\begin{table}
\caption{\label{tab:t2}$UBVI$ photometric observations of standard
stars.}
\centering
\begin{tabular}{lcccc}
\hline\hline
Target&Date&Filter&Exposure(s)&airmass\\
\hline

G93-48&N1&$U$&2$\times$200&1.34~-~2.00\\
&&$B$&3$\times$60&1.18~-~2.14\\
&&$V$&2$\times$40&2.08~-~2.10\\
&&$I$&3$\times$30&1.18~-~1.91\\
&N2&$U$&3$\times$200&1.17~-~2.01\\
&&$B$&2$\times$60,40&1.17~-~2.07\\
&&$V$&3$\times$40&1.17~-~2.12\\
&&$I$&3$\times$30&1.17~-~1.92\\
\hline
PG2213-006A&N1&$U$&3$\times$200&1.14~-~1.86\\
&&$B$&3$\times$60&1.15~-~1.74\\
&&$V$&3$\times$40&1.14~-~1.77\\
&&$I$&3$\times$30&1.14~-~1.92\\
&N2&$U$&4$\times$200&1.14~-~2.00\\
&&$B$&4$\times$60&1.14~-~1.88\\
&&$V$&4$\times$40&1.14~-~1.92\\
&&$I$&4$\times$30&1.14~-~2.06\\
\hline
TPHE&N1&$U$&60,100&1.07~-~1.08\\
&&$B$&40,80&1.10~-~1.10\\
&&$V$&30,60&1.09~-~1.09\\
&&$I$&10,30&1.08~-~1.08\\
&N2&$U$&200&1.06~-~1.06\\
&&$B$&20,60,80&1.06~-~1.07\\
&&$V$&5,20,60&1.07~-~1.07\\
&&$I$&5,10,30&1.06~-~1.06\\

\hline
\end{tabular}
\end{table}

Basic calibration of the scientific CCD frames was done using IRAF
package CCDRED. For this purpose, zero exposure frames and
twilight sky flats were taken every night. All frames were pre reduced,
applying trimming, bias, and flat-field correction. Before
flat-fielding, all frames were corrected for linearity, following
the recipe discussed in \citet{Ham2006}.
Photometry was then performed using the IRAF
DAOPHOT/ALLSTAR and PHOTCAL packages. Instrumental
magnitudes were extracted following the point-spread function (PSF) method \citep{ste1987}. Aperture corrections were
then determined, making aperture photometry of a suitable
number of well isolated stars (typically 20 to 90) all above the target fields. Typically these corrections were found to vary from 0.14 to 0.31 mag,
depending on the filter. The PSF photometry was finally aperture corrected, filter by filter.\\

\noindent
The transformation equations to put instrumental magnitudes into the $UBVI$ standard system were of the
form:

\begin{center}
\noindent
\hspace*{0.5cm} $u=U+u_1+u_2 \times X+u_3 \times (U-B)$ \\
\hspace*{0.5cm} $b=B+b_1+b_2 \times X+b_3 \times (B-V)$ \\
\hspace*{0.5cm} $v=V+v_1+v_2 \times X+v_3 \times (B-V)$ \\
\hspace*{0.5cm} $i=I+i_1+i_2 \times X+i_3 \times (V-I)$ \smallskip
\end{center}

\noindent where $u_2$, $b_2$, $v_2$ y $i_2$ are the extinction
coefficients for the $UBVI$ bands, $X$ is the air mass for each
exposure and $u_1$, $b_1$, $v_1$, $i_1$, $u_3$, $b_3$, $v_3$, and
$i_3$ the fitted parameters. As said, extinction coefficients were
computed for each night. Final photometric tables will be made available at CDS.

\begin{table}
\caption{\label{tab:t3} Extinction coefficients}
\centering
 \begin{tabular}{lcccc}

\hline
\hline
$Night$ & $u_1$ & $u_3$ & $b_1$ & $b_3$\\
\hline
N1&3.89$\pm$ 0.04&-0.29$\pm$ 0.01&1.82$\pm$ 0.01&-0.08$\pm$ 0.01\\
N2&3.83$\pm$ 0.02&-0.30$\pm$ 0.01&1.80$\pm$ 0.01&-0.08$\pm$ 0.01\\

\hline
&$v_1$&$v_3$&$i_1$&$i_3$\\
\hline
N1&1.71$\pm$ 0.01&0.08$\pm$ 0.01&1.81$\pm$ 0.01&-0.05$\pm$ 0.01\\
N2&1.73$\pm$ 0.01&0.08$\pm$ 0.01&1.68$\pm$ 0.02&-0.04$\pm$ 0.01\\ 

\hline
& $u_2$ & $b_2$ & $v_2$ & $i_2$\\ 
\hline
N1 & 0.37$\pm$0.02 & 0.21$\pm$0.01 & 0.13$\pm$0.01 & 0.03$\pm$0.01\\
N2 & 0.41$\pm$0.02 & 0.22$\pm$0.01 & 0.14$\pm$0.01 & 0.03$\pm$0.01\\           
\hline
\hline
\label{coe1}
\end{tabular}
\end{table}

\noindent
A comparison of our photometric results was performed at the end of the process. We found 60 stars in common with \citet{Heske,Heske1985} but seven of them are probably variable stars and were rejected. The remaining stars were then used to compute mean differences and standard deviations in the sense our measures minus theirs. The procedure yields: \\

\noindent
$\Delta V=0.012 \pm0.055$,\\
\noindent
$\Delta (B-V)=-0.006\pm0.047$,  and \\
\noindent
$\Delta (U-B)=0.145\pm0.085$.\\

\noindent
Finally,
the transformation from pixels (i.e., detector-coordinates) to equatorial right ascension and declination for the equinox J2000.0 was performed using Gaia DR2 \citep{Gaia} coordinates of 98 brightest stars in the field. The number of stars detected in the V and B filters amount at 21196. Among them 8719 stars have counterparts in the U filter and 20960 in the I filter. The completeness limits for each filters are $U_{lim} = 18.7$,  $B_{lim} = 19.9$,  $V_{lim} = 18.2$, and $I_{lim} = 20.2$.

\noindent
 The resulting catalog was cross-correlated with Gaia DR2 \citep{Gaia} data with 1.5" radius of cross-correlation (for 93 per~cent of objects angular distance between sources turned out to be less then 0.4"). Taking into consideration parallaxes and using isochrones \citet{Marigo} we calculated an approximate mass limit of  0.85-1.0 $M_{\sun}$. The number of matches with Gaia was 21172 objects, but only objects with errors in parallaxes being less than 20$\%$ were kept for the following analysis, so the number of objects in the final catalogue tuned out to be 11506 objects.

	\section{Data analysis, physical groups and their properties}
	
	In this Section we describe how stellar groups are identified, and how individual membership has been assessed. Then, properties of the various groups are derived.

	\subsection{The clustering algorithm}
	
	We applied a clustering algorithm in a 5-dimensional space using equatorial coordinates, proper motion components and parallaxes. We should mention that in the framework of clustering methods the term {\it cluster} does not refer to a physically bound star cluster but to a set of objects sharing common properties and clustering in a N-dimensional coordinate space. We will call the outcome of the clustering algorithm a {\it cluster} or a {\it group}, and use the term {\it star cluster} for physical astronomical objects.\\

\noindent	
	The clustering algorithm here adopted is based on the DBSCAN (Density-Based Spatial Clustering of Applications with Noise) technique. The algorithm was implemented in Python language and uses clustering module of machine learning library \textsc{scikit-learn} \citep{scikit-learn} as its basis. One of the main features of DBSCAN is that it considers clusters as a set of core points in the neighbourhood of each other, non-core points in the neighbourhood of core points and noise. So, in comparison with other commonly used methods, such as k-means and affinity propagation methods, DBSCAN has an important advantage, namely that part of the points are considered to be noise, non-members of clusters. This feature was one of the crucial points when choosing DBSCAN as the best method for our purposes.
	
	DBSCAN requires numeric values of two parameters: (1) {\it  eps} - maximum distance between two points to label that one is in the neighbourhood of the other and (2) {\textit N} - number of points in a neighbourhood of a point to label it as a core point. The choice of the parameters was performed the following way:
	
	\begin{description}
	\item (a) Testing parameters from wide intervals: (0.01-0.99) and (10-60) for the first ({\it eps}) and the second ({\it N}) parameter, correspondingly. Lower limit for {\it N} was chosen in agreement with \citet{Schubert}, where it is recommended to adopt {\it N} values which are not less than twice the value of dimensions of the data. As in the present work clustering is performed in 5-dimensional space, the minimum value of {\it N} is 10.
	\item (b) Choosing parameters which lead to a maximum number of clusters: {\it eps}=0.17, {\it N}=10.
        \end{description}

	In order to get probability of each star to be a cluster member and to take into account errors of observables, the clustering algorithm was applied 1000 times. In each run data was altered by adding random value extracted from a normal distribution with the mean value equals to original data and dispersion equals to error. Errors were taken individually, so for each star parameters were taken from individual distributions. The average mean values of proper motions errors are $e_{\mu_{\alpha}*} = 0.25 $ mas\,year$^{-1}$, $  e_{\mu_{\delta}} = 0.17$ mas\,year$^{-1}$ and mean error of parallaxes are $e_{plx} = 0.13$ mas. Membership probability was then determined according to the number of events when a star was labeled as a member of a cluster divided by the number of runs (1000).\\

\noindent	
	Before the application of the above mentioned algorithm, data were transformed using principal component analysis (PCA) to eliminate correlations between observables. PCA constructs an orthogonal coordinate system in such a way that all the subsequent principal components have the largest variance and are orthogonal to the previous ones: the first principal component has the largest variance, the second has the largest variance and orthogonal to the first one, and so forth. Then all principal components were scaled to unit variance.
	
	\begin{figure}
		\includegraphics[width=\columnwidth]{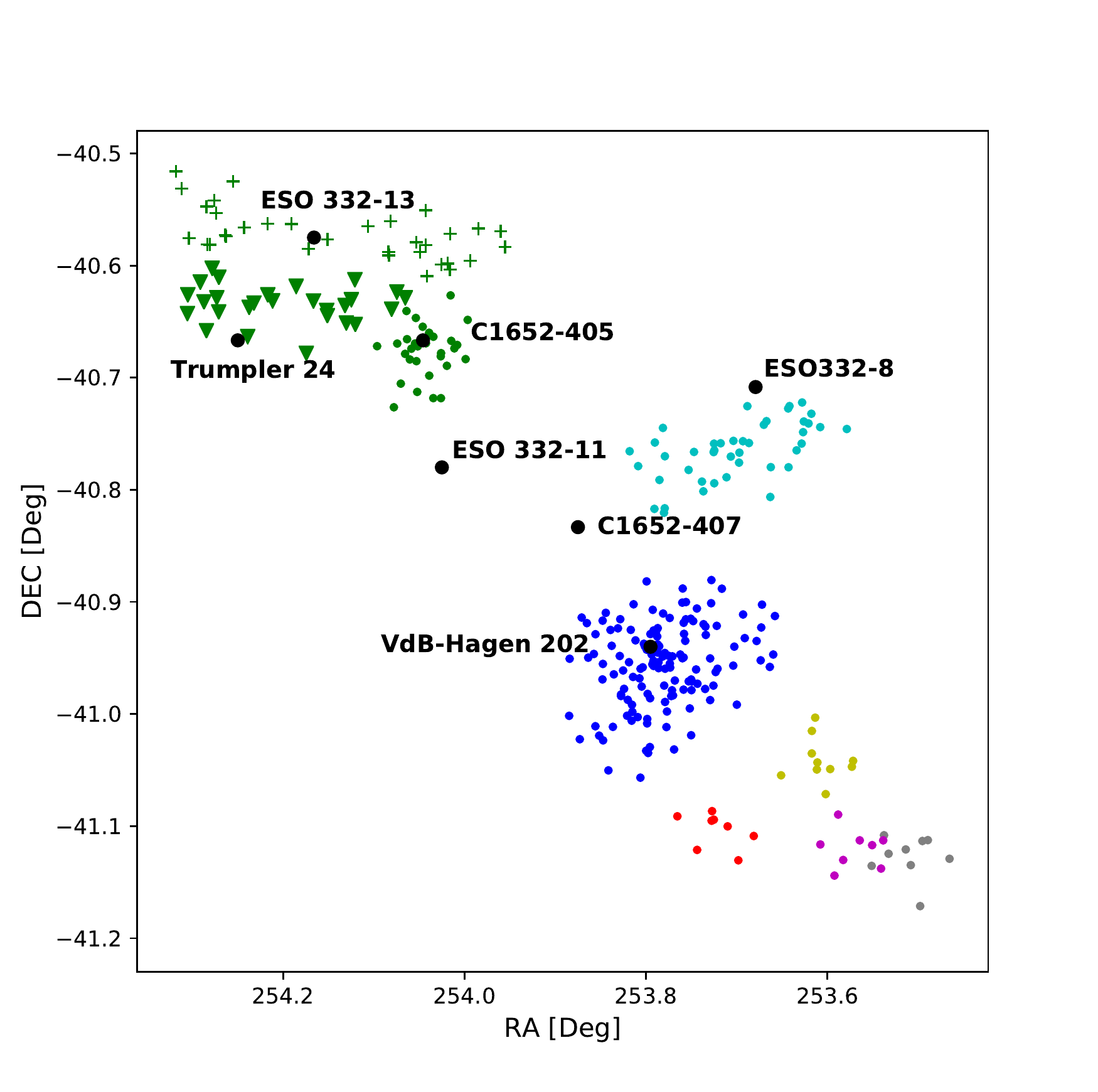}
		\caption{Detected groups in the covered Sco OB1 association region and comparison with known or suggested groups from Simbad. See text for details.}
		\label{fig:Fig3}
	\end{figure}
	
	As a result of the clustering algorithm we obtained seven groups with stars having membership probability of more than 50 per cent (Fig.~\ref{fig:Fig3}). For members of groups B (which we subdivided into B1, B2, and B3 subgroups) and A the probability is larger than 90 per cent and 98 per cent,  correspondingly, according to probability distribution peaks. 
	For all of these groups we estimated fundamental properties: age, reddening, distance and mean proper motion. 
	Distances were obtained by two independent ways: with the use of photometry alone,  and from Gaia DR2 parallaxes.
	
	\begin{figure}
		\includegraphics[width=\columnwidth]{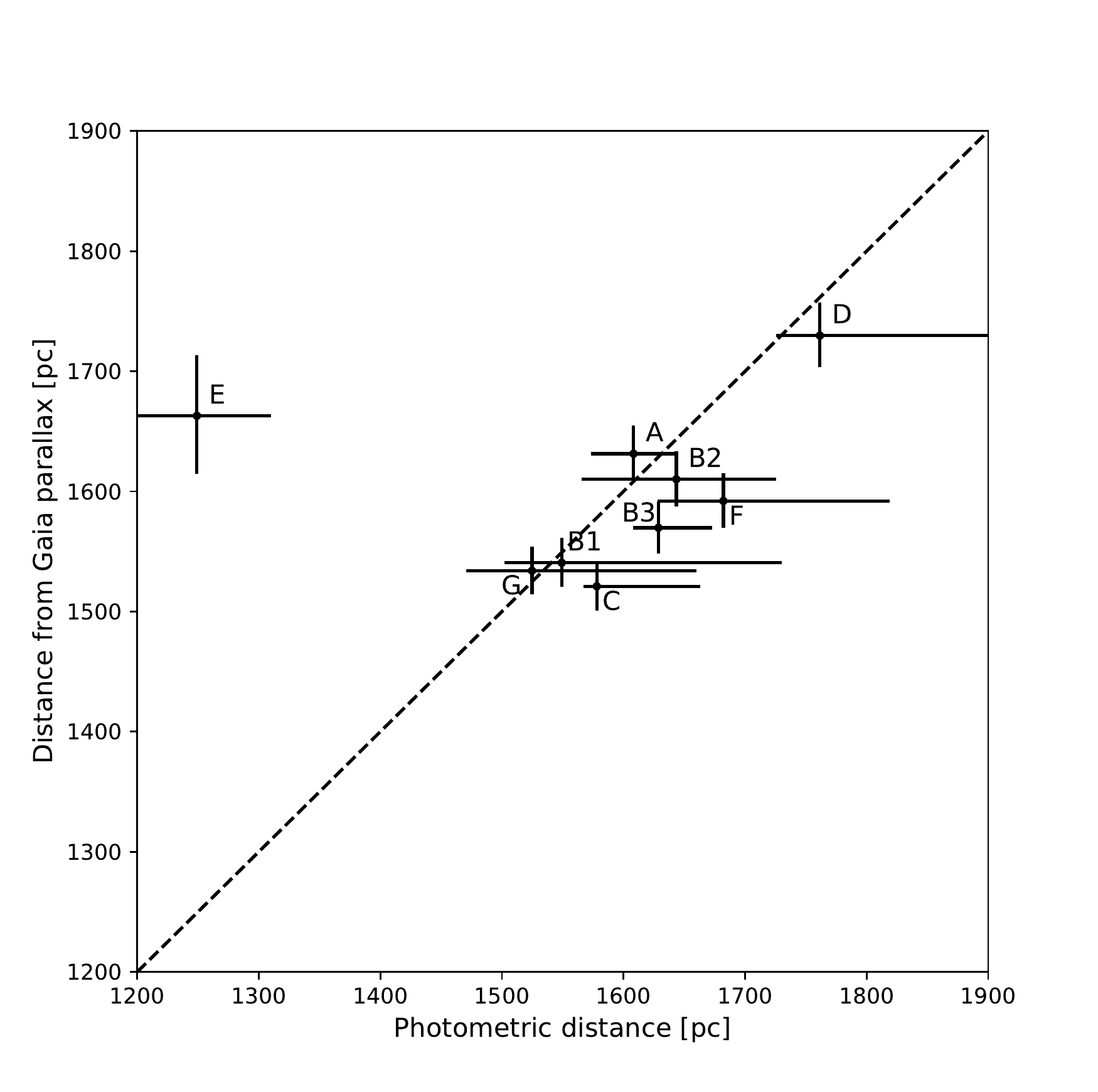}
		\caption{Photometric distances vs. distances from Gaia parallaxes}
		\label{fig:Fig4}
	\end{figure}

	\begin{figure}
		\includegraphics[width=\columnwidth]{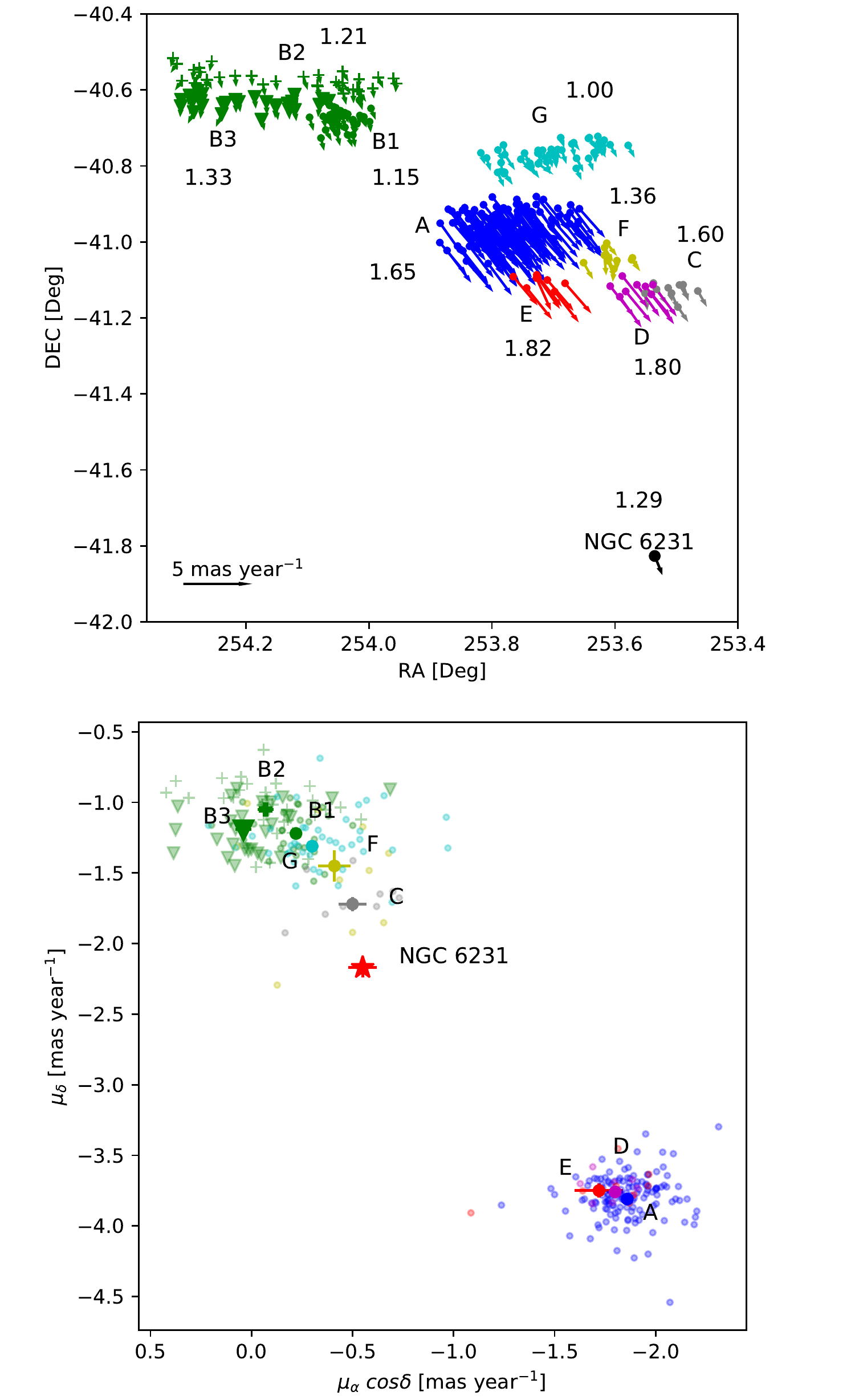}
		\caption{Upper panel: tangential motion of the various groups. The sizes of the arrows are proportional to the velocities. We also indicate the tangential motion of NGC 6231 \citep{Kuhn} for the sake of comparison. Lower panel: vector point diagram. Colours indicate the various groups, as in Fig.~3, while numbers report the interstellar absorption from Table~4.}
		\label{fig:Fig5}
	\end{figure}
	
	\subsection{Photometric distances}
	
	We adopted a zero-age main sequence (ZAMS) from \citet{Turner} to estimate reddening and distance modulus by fitting theoretical ZAMS to our data in $(B-V)$ vs. $(U-B)$ colour-colour and $(B-V)$ vs. $V$ colour-magnitude diagrams respectively (Fig.~\ref{A1:ccd} and Fig.~\ref{A1:cmd} in Appendix).
	The region of Sco OB1 association is highly affected by variable reddening \citep{Damiani} and probably affected too by differential extinction. 
	To obtain more precise results we derived inclinations of reddening vectors and ratios of total-to-selective absorption $R_V=A_V/E(B-V)$ in the direction of each group. We performed a weighted least squares fit in $(V-I)$ vs. $(B-V)$ colour-colour diagram, where stars are distributed almost parallel to the reddening vector. Then we used a parametrisation for optical/NIR wavelengths regions as described in  \citet{Cardelli} and obtained values of $R_V$ and $(U-B)/(B-V)$. The resulting values of $R_V$ toward the direction of each group are listed in Table~\ref{tab:t4}. Reddening E(B-V) is then derived from the colour-colour diagram in a standard way \citep{carraro}.
	
	\begin{table*}
		\caption{Fundamental parameters of the detected groups. $R_V$ indicates the ratio of total-to-selective absorption in the direction of each group.}
		\label{tab:t4}
		\begin{center}
			\begin{tabular}{c c c c c c c c c}
				\hline
				Group    & $E(B-V)$  & $\sigma_{E(B-V)} $ &  Distance                &   log(Age)& $\sigma_{log(Age)}$  & $R_V $ & $\sigma_{R_{V}} $ &  $A_{V}$  \\ 
				& mag       & mag                 & pc                       &    dex   &  dex  &        &                   &           \\ 
				\hline                                                                                                                    
				A        & 0.55       &  0.10               & $ 1608_{-35}^{+36}   $   &   8.75   &  0.10  & 3.0    &  0.2     &   1.65    \\ 
				B1       & 0.36       &  0.10               & $ 1549_{-47}^{+181}  $   &   6.45   &  0.30  & 3.2    &  0.2     &   1.15   \\ 
				B2       & 0.39       &  0.30               & $ 1644_{-78}^{+82 }  $   &   6.75   &  0.40  & 3.1    &  0.2     &   1.21   \\ 
				B3       & 0.38       &  0.20               & $ 1629_{-21}^{+44 }  $   &   6.95   &  0.30  & 3.5    &  0.2     &   1.33    \\ 
				C        & 0.57       &  0.20              & $ 1578_{-11}^{+85 }  $   &   6.85   &  0.20  & 2.8    &  0.3    &   1.60   \\ 
				D        & 0.58       &  0.20               & $ 1761_{-36}^{+139}  $   &   8.00    &  -    & 3.1    &  0.3    &   1.80   \\ 
				E        & 0.57       &  0.20               & $ 1249_{-58}^{+61 }  $   &   8.15   &  -    & 3.2    &  0.3   &   1.82   \\ 
				F        & 0.47       &  0.10               & $ 1682_{-54}^{+137}  $   &  $>6.60$  &  0.50  & 2.9    &  0.2     &   1.36   \\ 
				G        & 0.40       &  0.20               & $ 1524_{-54}^{+135}  $   &   6.70    &  0.30  & 2.5    &  0.2     &   1.00    \\ 
				
				\hline
			\end{tabular}
		\end{center}
	\end{table*}

	\subsection{Astrometric distances}
	
	Different estimations of the zero-point offset of Gaia parallaxes have been recently found. In the paper by \citet{Lindegren} a claim is made for a zero-point offset of 0.03~mas, but there are evidences that it is larger. \citet{Zinn} estimated a value of about 0.053~mas using red giant branch stars from the APOKASC-2 catalog. \citet{Riess} obtained a value of 0.046~$\pm$~13~mas using bright extragalactic Cepheids. All of them indicate that Gaia parallaxes are underestimated. Here we used the value of offset of 0.045~$\pm$~0.009~mas \citep{Yalyalieva} obtained from comparison of parallaxes from photometry study of young open clusters in the  northern sky with their parallaxes from Gaia. 
	We applied kernel density estimator to parallax distribution. We used Gaussian as a kernel with the optimal bandwidth found from cross-validation algorithm individually for each cluster.  To take into account individual errors of parallaxes, we added weights to each point equals to $1/e_{plx}^2$, where $e_{plx}$ - error of parallax of each star. Although for our analysis we selected stars with errors in parallaxes less then 20$\%$, errors for stars belonging to groups are much less - mean error of parallaxes ranges form 6$\%$ for C group to 11$\%$or D group.
		 The outcome of kernel density estimation was fitted with Gaussian distribution, then stars out of $2\sigma$ were removed. This $\sigma$-clipping procedure was repeated until convergence. When the number of stars in the group was not enough for statistics, the mean value was adopted. To take into account the  error of zero-point offset we shifted parallaxes to $\pm$0.009~mas and repeated Gaussian fitting in the way mentioned above. The resulting estimated error of mean distance to group is considered to be interval between mean values of shifted parallaxes with added errors from the Gaussian fit.\\
\noindent
A comparison between photometric and astrometric distances is shown in Fig.~\ref{fig:Fig4}. There is a generally good agreement between the two different estimates, except for group E. 
Because of the global good agreement we will adopt in the following the astrometric distance for group E as well. This is supported also by the fact that group E shares all the other properties (age and kinematics) with groups A and F, and also because the photometric distance estimate could be affected by low number statistics.

	\subsection{Ages}
	
	Ages for groups A, B, C, F and G were estimated by fitting theoretical PARSEC + COLIBRI isochrones \citep{Marigo} on the data in the colour-magnitude diagram $(B-V)$ vs. $V$ (see Appendix \ref{A1:cmd}). Isochrones were previously shifted by the values of colour excess $E(B-V)$ and distance modulus $(m-M)_V$, obtained from ZAMS fitting. Resulting ages are listed in the Table~\ref{tab:t4}.\\
	
\noindent	
Group A: about 60$\%$ of the stars are non-ZAMS, with 5 evolved stars in the red clump around $(B-V)\approx 1.7$ and $V \approx 13$. ZAMS stars are the stars that are fainter than $V\approx15.5$ and redder than $(B-V)\approx 0.9$.
	
Group B1: group of 29 young stars, about 75$\%$ of them are PMS stars, ZAMS stars could be found in the color range $0.15 \loa (B-V) \loa 0.3$ and with magnitudes $11.5 \loa V \loa 12.8$.
	
Group B2: about 70$\%$ of 32 stars are PMS stars with some spread in ages, ZAMS stars have colours about $0.2 \loa (B-V) \loa 0.4$ and magnitudes in $V$ filter about $12 \loa V \loa 14$.
	
Group B3: like in the cases of B1 and B2 groups, the number of ZAMS stars aisabout 25-30$\%$, and  they have colours $0.2 \loa (B-V) \loa 0.5$ and magnitudes $11.2 \loa V \loa 14.3$. The remaining 70-75$\%$ of 27 stars are PMS stars.
	
Group C: 5 out of 9 stars are supposed to be ZAMS stars in the range of $0.4 \loa (B-V) \loa 0.6$ and $12.5 \loa V \loa 14$, and 4 stars are PMS.
	
Group F: a small group of 10 stars, 3 of them are on ZAMS ($0.3 \loa (B-V) \loa 0.45$, $11.7 \loa V \loa 13.4$). 7 PMS stars are scattered, causing some ambiguity in ages.
	
Group G: a group numbers 41 objects, about 34$\%$  are on ZAMS in the following ranges: $0.2 \loa (B-V) \loa 0.55$ and $11 \loa V \loa 13.9$

\noindent
Groups D and E are not rich enough in bright stars and isochrones fitting could lead to inaccurate results. In these two cases we estimated ages of the brightest stars still in MS and adopted them as ages of the groups.

	\subsection{Mean proper motions}
	
	Mean proper motion components were estimated via a $\sigma$-clipping method, and removing stars beyond $3\sigma$ threshold. They are shown in the vector point diagram (lower panel of Fig.~\ref{fig:Fig5}). Proper motion components seem to form two distinct groups. We also computed the tangential velocities, which are shown in the upper panel of the same Figure, where we  indicate both the direction and the size of the proper motion vectors for  each star in the various groups. The solid black arrow in the figure indicates the scale of tangential velocities. Also in this panel two separate groups are visible.
Mean proper motion components and tangential velocities of groups are   listed in Table~\ref{tab:t5}. 	

	
	\begin{table*}
		\caption{Physical parameters of groups from Gaia data}
		\label{tab:t5}
		\begin{center}
			\begin{tabular}{c c c c c c c c c c}
				\hline
				Group &  $\alpha$  &  $\delta$ & Distance &  $ \mu_{\alpha}*$  &  $\sigma_{\mu_{\alpha}*}$ &  $ \mu_{\delta} $  &  $\sigma_{\mu_{\delta}}  $  & $v_T$ & $\sigma_{v_T}$ \\
				&    deg       &   deg     & pc                   & mas\,year$^{-1}$ &  mas\,year$^{-1}$       &   mas\,year$^{-1}$  & mas\,year$^{-1}$  &  km\,s$^{-1}$ &  km\,s$^{-1}$  \\
				\hline                           
			A     &   253.781     & -40.956  &  $1631_{-22}^{+23}$ & -1.86        & 0.01                 &   -3.81         & 0.01 &  33.46 & 0.81   \\
			B1    &   254.042     & -40.677  &  $1540_{-19}^{+20}$ & -0.22        & 0.03                 &   -1.22         & 0.03 &  9.25 & 0.59    \\
			B2    &   254.141     & -40.571  &  $1610_{-22}^{+23}$ & -0.07        & 0.04                 &   -1.05         & 0.03 &  8.11 & 0.29   \\
			B3    &   254.199     & -40.635  &  $1569_{-21}^{+21}$ &  0.04        & 0.04                 &   -1.19         & 0.03 &  8.95 & 0.5   \\
			C     &   253.510     & -41.128  &  $1521_{-20}^{+20}$ & -0.50        & 0.07                 &   -1.72         & 0.05 &  13.08 & 0.55   \\
			D     &   253.571     & -41.120  &  $1729_{-26}^{+27}$ & -1.80        & 0.04                 &   -3.76         & 0.03 &  34.07 & 0.8   \\
			E     &   253.722     & -41.103  &  $1663_{-48}^{+50} $ & -1.72        & 0.12                 &   -3.75         & 0.05 &  32.54 & 1.06   \\
			F     &   253.606     & -41.041  &  $1592_{-22}^{+23}$ & -0.41        & 0.08                 &   -1.45         & 0.11 &  11.42 & 0.88    \\
			G     &   253.701     & -40.765  &  $1533_{-19}^{+20}$ & -0.30        & 0.03                 &   -1.31         & 0.02 &  9.82 & 0.49   \\

				\hline
			\end{tabular}
		\end{center}
	\end{table*}
	
	\subsection{Group by group properties}
	
	In the Table~\ref{tab:t4} and Table~\ref{tab:t5} we presented parameters of groups estimated from photometry and from Gaia data, respectively. For most groups only stars with more than 50 per cent probability were kept, except for  the richest A and B groups where the threshold was enhanced to 98 per cent and 90 per cent respectively in accordance with peaks of the probability distributions. Table~\ref{tab:t4} also includes (last two columns) the tangential velocity $v_T$ and the associated uncertainty. \\

\noindent	
To further confirm their nature,
we searched for previously detected groups in the area using SIMBAD and we
compared their positions with the groups we identified. Some of them have compatible coordinates. VdB-Hagen 202 is very close to  group A, while the three clusters C 1652-405, ESO 332-13 and Trumpler 24 cover the area of our group B (B1, B2, and B3). ESO 332-8, on the other hand, lies not far from our group G.  Two presumed clusters from SIMBAD, ESO 332-11 and C1652-407 do not have any counterpart in our study.  We suggest these two latter are most probably spurious detections. 

Finally, our groups C, D, E, and F are not present in SIMBAD.
In Table~\ref{tab:t6} we listed coordinates of the SIMBAD objects and some of their physical properties (distance, age, proper motion) as taken from \citet{Dias}.\\
		\begin{table*}
		\caption{Coordinates of objects and labels of the groups/clusters in the region according to SIMBAD and their properties according to \citet{Dias}.}
		\label{tab:t6}
		\begin{center}
			\begin{tabular}{c c c c c c  c c c c}
				\hline
			Cluster      &   Nearest group     & $\alpha$  &  $\delta$ & Distance & log(Age)
			    & $ \mu_{\alpha}*$, &  $\sigma_{\mu_{\alpha}*}$,&  $ \mu_{\delta}$,&  $\sigma_{\mu_{\delta}}  $\\
			&         &   deg      &   deg      & pc           &  dex      & mas\,year$^{-1}$ &  mas\,year$^{-1}$       &  mas\,year$^{-1}$   & mas\,year$^{-1}$         \\
			\hline                
			C 1652-405   &    B1    &    254.046  &  -40.667   &   2160 &   7.12   &   -1.72 & 0.18 &   -2.44 &  0.08 \\
			ESO 332-11   &        &    254.025  &  -40.780   &   1841 &   7.10   &   -5.28 & 0.60 &    0.86 &  0.53 \\
			ESO 332-13   &    B2    &    254.166  &  -40.575   &   2910 &   6.82   &   -2.73 & 0.05 &   -0.57 &  0.02 \\
			C 1652-407   &        &    253.875  &  -40.833   &   1000 &   -      &   -0.89 & 0.54 &   -4.33 &  0.61 \\
			Trumpler 24  &        &    254.250  &  -40.667   &   1138 &   6.92   &   -4.36 & 0.03 &   -0.17 &  0.04 \\
			ESO 332-8    &    G    &    253.680  &  -40.708   &   1200 &   8.17   &   -3.87 & 0.12 &   -0.52 &  0.16 \\
			VdB-Hagen 202 &   A         &    253.795  &  -40.940   &   1607 &   8.05   &   -2.55 & 0.21 &   -2.89 &  0.33 \\
			\hline
			\end{tabular}
		\end{center}
	\end{table*}

\noindent	
	\textbf{Group A}\\	
Group A  clearly coincides with VdB-Hagen 202 \citep{vdb}, who describe it as a poor, possibly embedded cluster.  It also corresponds to \citep{sege} group Trumpler 24 I.
It is  the richest group detected in the area.  It has also the largest tangential velocity and the largest age.
The isochrone fit yields a 500 million year, while both the astrometric and the photometric analysis support a heliocentric  distance of 1.65 kpc. This group has not been detected by
\citep{Damiani}, possibly because it does not contain young M stars. On the other hand, close to this position \citep{Kuhn} found two rich groups (3 and 5, according to their numbering) 
of young stars slightly to the north of our group A.
In any case, the large age and tangential motion lends additional support to a picture in which this cluster does not belong to the association, but it is probably caught in the act of passing through it.  The groups D and E (see below) share the same properties of this group A.

\noindent	
	\textbf{Group B}\\
This group is situated in the southern edge of the \ion{H}{ii} region G345-+1.50, and appears quite scattered. DBSCAN returns three density peaks, that we indicate as B1, B2, and B3, but the area roughly corresponds to Trumpler 24.  \citep{sege} also identified this group, which he indicated as 
group Trumpler 24 III. It is separated by a gap from the other groups identified in this study. 
Beside the location, these 3 groups share the same age and kinematics.
In the literature, Trumpler 24 distance ranges from 1.6 to 2.2 kpc \citep{sege,Heske}. Our study favours the shortest distance, both from photometry and from Gaia DR2 parallaxes. This group has clear counterparts both in \citep{Kuhn} and in \citep {Damiani}.

\noindent	
	\textbf{Group C}\\
	This is a very poor group located in the south west corner  of the field we covered. It is essentially composed by early type stars (early B spectral type, judging from the colour-colour diagram), and therefore it shares the same age as Trumpler 24 (group B).  The kinematic data support this association.

\noindent	
	\textbf{Group D}\\	
In spite of its vicinity to the previous group C, this group appears to be totally different. It does not contain young stars, and its kinematics is closer to VdB-Hagen 202 than to Trumpler 24. It seems also to be located somewhat in the background with respect to Trumpler 24. We suggest that this group is a part of the group A (see Table 5).
	
\noindent
	\textbf{Group E}\\
If we adopt the astrometric distance, this poorly populated group shares the same kinematics as group D, and lies very close to VdB-Hagen 202. Its paucity of stars prevents us from computing precise age. However, it seems plausible to adopt an age close to group D. We suggest that this group is a part of the group A (see Table 5).
	
\noindent
	\textbf{Group F}\\
This group shares the same properties of group C.  It coincides with \citep{Kuhn} group 3. A few early type stars, and age, kinematics and distance consistent with Trumpler~24.
	
\noindent
	\textbf{Group G}\\
	This corresponds to \cite{sege} group Trumpler 24 II. Similar to the C and F groups, this group is young. Even if it is situated right to the north of the old A group (VdB-Hagen 2020), it possesses mean proper motion components close to Trumpler~24.\\

\noindent
	Overall, this analysis of the detected group leads us to separate them in two families:
	
	\begin{description}
	
	\item {\it family I:} B, C, F and G groups have colour-colour diagrams typical of a very young population,  and show the presence of pre Main Sequence stars in the colour-magnitude diagram. They share similar proper motion components : $\mu_{\alpha}*=\mu_{\alpha}~\cos{\delta}= - 0.3$\,mas\,year$^{-1}$, $\mu_{\delta}= -1.3$\,mas\,year$^{-1}$.\\
	
	\item {\it family II:} A, D and (maybe) E groups are significantly older and  have compatible proper motion components: $\mu_{\alpha}*= - 1.7$\,mas\,year$^{-1}$, $\mu_{\delta}= - 3.7$\,mas\,year$^{-1}$. They have distances on average larger than {\it familiy I}.
	
	\end{description}

	\begin{figure*}
		\includegraphics[scale=0.5]{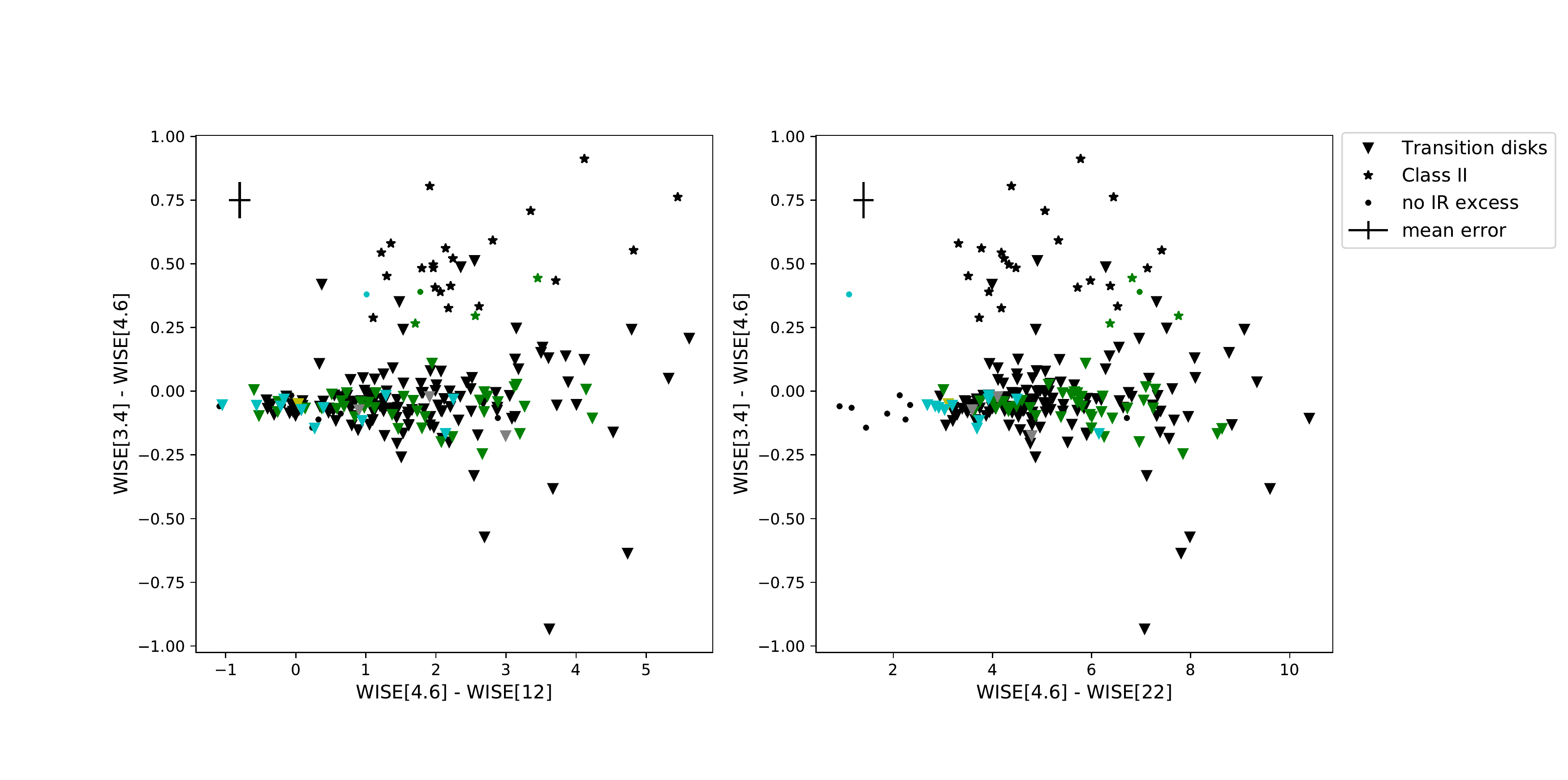}
		\caption{WISE two colour diagrams of detected PMS in the surveyed area. Colours refer to detected groups as in previous figures, while in black we refer to stars not associated with detected groups.}
		\label{fig:Fig6}
	\end{figure*}
	
\section{Pre main sequence stars}
\citep{Eggen76} firstly suggested that a violent episode may have taken place in Sco OB1 association, which rich in pre Main Sequence (PMS) stars. \citep{Heske} found a large number of PMS star candidates in some portions of ScoOB1 in the V range from 11 to 14 mag.  More recently, \citep{Damiani} performed an extensive search  of such stars in X-rays, in $H_{\alpha}$, and in  infrared.\\
\noindent
Our study  using deeper photometry, confirms the presence of these objects.
In fact the CMDs of groups B, C, F and G  (see appendix B) indicate the clear presence of stars in the region where PMS are expected to be.
These star are members of the groups according to our criteria. To further check their nature we make use of photometry in mid-infrared from WISE \citep{Wise}, and cross-correlate it with our data-set. It is well known that PMS stars which have not lost circum-stellar disks  completely exhibit an excess of emission in IR wavelengths ( $\lambda \geq 2 \mu$).  To probe this,
we closely followed the procedure described by \citet{Koenig}. PMS stars are divided in three classes:  Class III objects refer to PMS stars without disks (and also background stars, which we should not have in our sample), while Class II and I are young stellar objects with contracting envelopes and hence optically thick disks.  They are separated by different amount of infra-red excess.
Class I objects possess the highest infra-red excess, while Class II objects are intermediate. 
The  separation  is performed using the colour-colour diagrams  $[4.6] -  [12]$ vs. $[3.4] - [4.6]$, where $[3.4], [4.6], [12]$ are the WISE \citep{Wise} filters W1, W2, W3,  correspondingly.  There are also young objects which do not have excess emission at these wavelengths, but only beyond 20 $\mu$m,  and they could be detected with the use of the colour-colour diagram $[4.6] - [22]$ vs. $[3.4] - [4.6]$, where $[22]$ refers to W4 WISE filter. These objects are supposed to be so-called {\it transition disks} (TD), which may be the result of planet formation, photo-evaporation, or other processes that led to the disappearance of the inner disk.\\

\begin{figure}
		\includegraphics[width=\columnwidth]{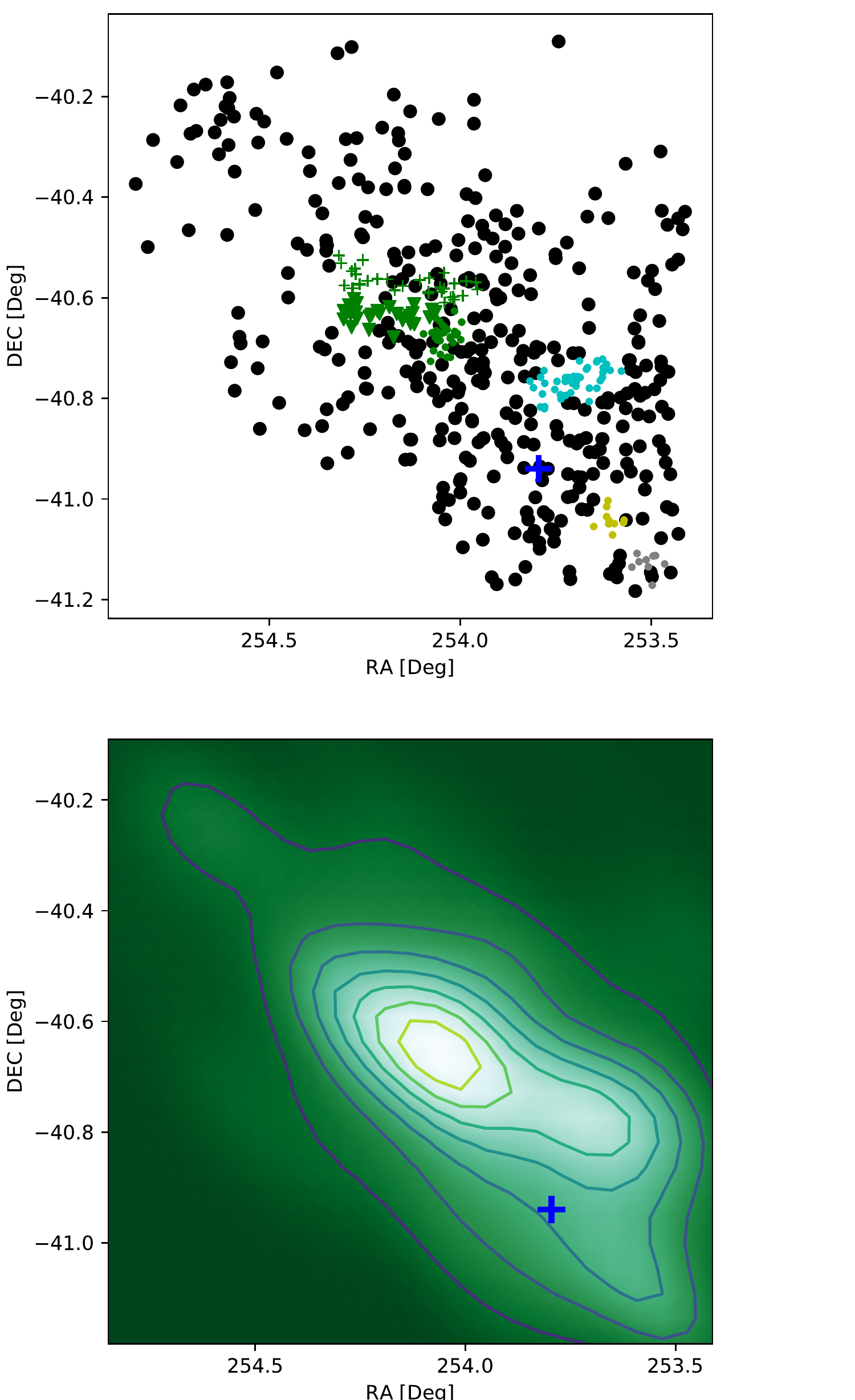}
		\caption{Spatial distribution (upper panel) and surface density map (lower panel) of stars belonging to {\it family I} in the covered area. Colour-coded are the various groups. In black we identified stars not associated with groups. The blue cross indicates the position of VdB-Hagen 202.}
		\label{fig:Fig7}
	\end{figure}

\noindent	
In detail, we extracted data from WISE for stars in groups B, C, F, and G adopting a  crossmatch radius of $2\arcsec$. We then considered only stars with photometry in filters W1, W2, W3 and W4. \\
The results are illustrated in Fig.~\ref{fig:Fig7},  and can be summarised as follow:\\

\noindent
\begin{description}
\item {\bf  subgroup B1}:  we identified 11 TD;  
\item {\bf subgroup B2} : we found 13 TD and 3 Class II objects;
\item {\bf subgroup B3} : 15 TD  are detected; 
\item {\bf group C}:  we identified 3 TD;  
\item {\bf group F} : only 1 TD  was found;
\item {\bf group G}: we identified 11 TD .
\end{description}

\noindent
Then, by using the same scanning criteria, we looked for PMS stars that are not associated with our groups of {\it family I}, but that share the same kinematics and distance. We found 158 stars (indicated in black in Fig.~\ref{fig:Fig6}) , 136 of which are TD, and 22 Class II objects. Their spatial distribution is shown in Fig.~\ref{fig:Fig7}. \\
Therefore, we found PMS stars everywhere in the area we have surveyed, down to V= 18 mag. This population of scattered PMS stars are therefore members of {\it family I}. The whole family  would then correspond to the very same Star Formation event whose outcome
appears to have a complicated spatial structure, with different degrees of concentration across the association.

As shown in Figs 10 and 11 in \citep{Damiani} the highest concentration of young objects of M-types in his work coincides with the locus occupied by our groups B, A, C, F, G and D (although group G is a bit displaced to the north-west). \\

\noindent
According to \citep{Damiani} the visual absorption $A_V$ in Sco OB1 steadily decreases moving from NGC 6231 toward Trumpler 24. If we adopt the widely accepted value $A_V$ = 1.29 (E(B-V) =0.43)  for NGC 6231 and include it together with the respective values determined for the seven groups in Table 4, the absorption increases from NGC 6231 to groups C, D, E, and A. Group F, slightly to the est of group A, shows an absorption value quite close to the one of NGC 6231. Immediately north of group A, group G appears showing the lowest absorption value ($A_V$= 1) in the surveyed region. The remaining three subgroups B1, B2, and B3 show, in turn, small $A_V$ values, close to the one of NGC 6231. Therefore we can confirm \citep{Damiani} result in the sense that Tr 24-S is more affected by visual absorption than Tr 24, but we notice as well that NGC 6231is less reddened than Tr24-S. \\

\noindent
 The spatial distribution of {\it family I} 
stars (both OB  and pre MS stars) is shown in Fig~\ref{fig:Fig7}. 
The density map in the lower panel shows how all PMS stars in {\it family I} increase in concentration toward Tr 24 (group B). The upper panel containing all PMS stars found in the area confirms the trend that has been also reported in Damiani (2018) this is, Halpha emission, UV excess and NIR emission stars increase in number from Tr 24-S toward Tr 24 and IC 4628 exactly as our PMS stars do. Following upper panel in Fig. 7, it is seen an evident downfall in the density number of PMS stars north east of Tr 24. However, density numbers in this figure should be taken with care: the region north of Tr 24 is dominated by the HII region G345.45+1.50 whose brightness prevent us to get deeper observations in our UBVI photometry. So, we see that despite the reddening is constantly decreasing toward the north part and therefore we could in principle detect more, fainter, stars, the brightness of the nebula becomes higher thus reducing this advantage. We are confident  this a reasonable argument to explain why many low mass stars (at larger magnitudes) may have not been detected in our passband.

\begin{figure}
		\includegraphics[width=\columnwidth]{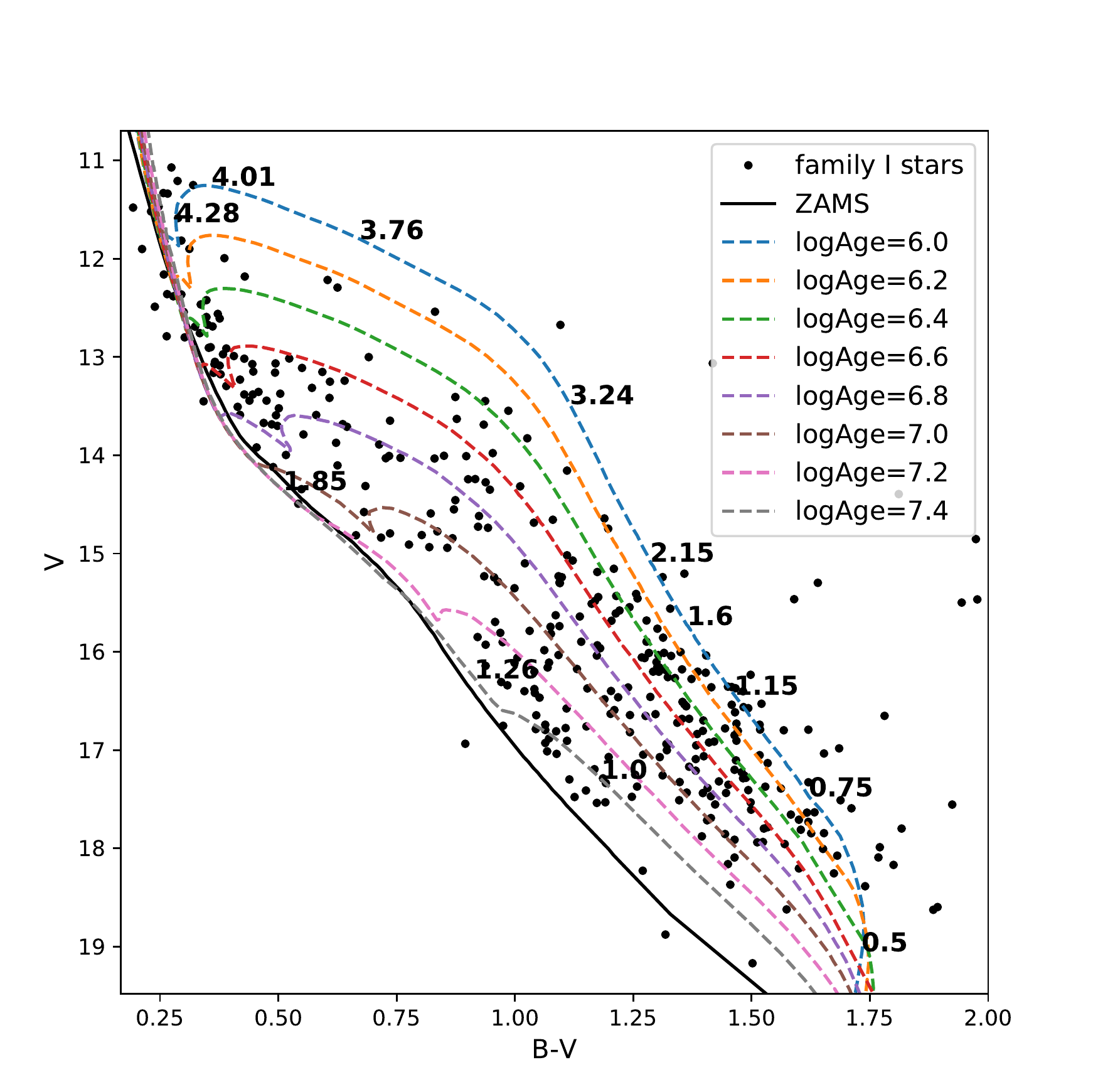}
		\caption{CMD of all the stars of {\it family I} we detected in the covered area. Super-imposed are PMS isochrones from \citep{Marigo}. A few values of star masses are indicated for illustration purposes.}
		\label{fig:Fig8}
	\end{figure}

\noindent
We finally draw our attention to the CMD of {\it family I} stars with a zoom on the PMS stars region, which is shown in Fig.~\ref{fig:Fig8}. Super-imposed are stellar isochrones from the Padova suite of models \citep{Marigo} for the labelled ages. We indicate also a few representative value of the stellar masses, as extracted from the isochrones. We can notice that PMS stars in this area of Sco OB1 have a mass range from  0.5 to 3 $M_{\odot}$.

	\section{Discussion and conclusions}
In this study we presented a new high quality data-set of optical photometry in the north east region of the Sco OB1 association. When combined with Gaia DR2 data, this data-set allowed us
to characterise the region in terms of stellar populations: age, distance, reddening, and kinematics.\\
The analysis of all this material led us to identify several groups, with different fundamental properties, which we divide into two families :

\begin{description}
	\item {\it family I:} B (B1, B2, and B3), C, F and G groups have colour-colour diagrams typical of a very young population: OB stars in the MS and PMS stars. They share similar proper motion components : $\mu_{\alpha}*= - 0.3$\,mas\,year$^{-1}$, $\mu_{\delta}= -1.3$\,mas\,year$^{-1}$. They also have compatible distances, being the average distance 1560$\pm$35 pc.\\
	Overall, B group coincides with Trumpler~24, while groups C, F and G are generally referred to as Trumpler 24-S \citep{Damiani}.
	We characterise also the PMS population both in these groups and in the general field. PMS stars sharing the same kinematics occupy the entire region we surveyed, although
	their density seems to increase northward, moving from Trumpler~24-S to Trumpler~24. This can be an effect of the decreasing amount of reddening.\\
	
	\item {\it family II:} A, D and  E groups are significantly older and  share compatible proper motion components: $\mu_{\alpha}*= - 1.7$\,mas\,year$^{-1}$, $\mu_{\delta}= - 3.7$\,mas\,year$^{-1}$. They have  on average distances larger than {\it family I}.  We confirms that group A (VdB-Hagen 202), which was in the past considered a dubious cluster,  is a physical object. It appears to be in an intermediate age open cluster, unexpected in this environment,  which happens to lie inside (and moving through) the association. Given the different tangential  velocity (both direction and value) the cluster is possibly caught in the act of crossing the association. Groups D and E, with similar properties, are most probably pieces of the same cluster. They can be the effect of VdB-Hagen 202 tidal dissolution. Another piece of evidence in this direction comes from the detection of a gap in the distribution of {\it family I} stars, as shown in Fig.~\ref{fig:Fig7}. The relation of VdB-Hagen 202 with the association can be better explored once radial velocities will be available that permit its orbit integration.
	
\end{description}
	
\noindent
The young rich open star cluster NGC 6231 is situated in the close vicinity of the area under investigation and has been a target of intense study in the past \citep{Reipurth}. The logAge is supposed to be about 6.3 - 6.9 dex \citep{Damiani2016,Sung}. As for distances, \citet{Reipurth} noted that the mean estimation of distance is 1.6 kpc. \citet{Dias} listed the distance of 1243 pc, which seems to be underestimated. \citet{Kuhn} found the distance of 1710 pc from Gaia DR2 parallaxes. We adopted here distance to NGC 6231 equals 1585 pc, given by \citet{Sung}, which is widely accepted \citep{Damiani}.

The age of NGC 6231 is in good agreement with the ages of {\it family I} groups. We adopted proper motion $\mu_{\alpha}= - 0.55 \pm 0.07$\,mas\,year$^{-1}$, $\mu_{\delta}= - 2.17 \pm 0.07$\,mas\,year$^{-1}$ from \citet{Kuhn} and plot it with the mean proper motion values of the groups (Fig.~\ref{fig:Fig5}). Again, its value is close to the estimations for the {\it family I}.

Thus, groups from sample I have roughly the same distances and age of  NGC 6231 and share simillar kinematics. They all are very young, approximately the same age as NGC 6231. These facts leads us to the idea of the connection of star formation histories of {\it family I} groups and NGC 6231.
	
In conclusion,  {\it family I} is a stellar population associated with the association whose centre is NGC 6231, while {\it family II} is an older population which is not related to the association.\\
With the third release of Gaia data, which is expected to be in the second half of 2021, it will be possible to get a better understanding of processes of this region. With the knowledge of radial velocities one will have the opportunity to trace orbits of objects of interest back in time, and estimate the most probable time and place of formation.

\section*{acknowledgements}
L. Yalyalieva acknowledges the financial support of the bilateral agreement between Lomonosov Moscow State University and Padova University, that allowed her to spend a period in Padova, where this work was completed. G. Carraro acknowledges the same agreement that allowed him to visit Lomonosov Moscow State University several times.
L. Yalyalieva was also supported by the Russian Foundation for Basic Research (projects 19-02-00611, 18-02-00890). E.Costa acknowledges support from the chilean Centro de Excelencia en Astrof\'isica y Tecnolog\'ias Afines (CATA) BASAL AFB-170002, and FONDECYT/CONICYT grants \#1110100 and \#1190038.
RAV and LR acknowledge the financial support from PIP 317 (CONICET) and to the Fac. de Ciencias Astron\'imicas y Geof\'isicas (UNLP).

	\bibliographystyle{mnras}
	\bibliography{references} 

	\cleardoublepage
	\appendix
	
	\section{Color-color diagrams}
	
	\begin{figure*}
		\begin{minipage}[h]{0.3\linewidth}	
			\center{\includegraphics[width=1\linewidth]{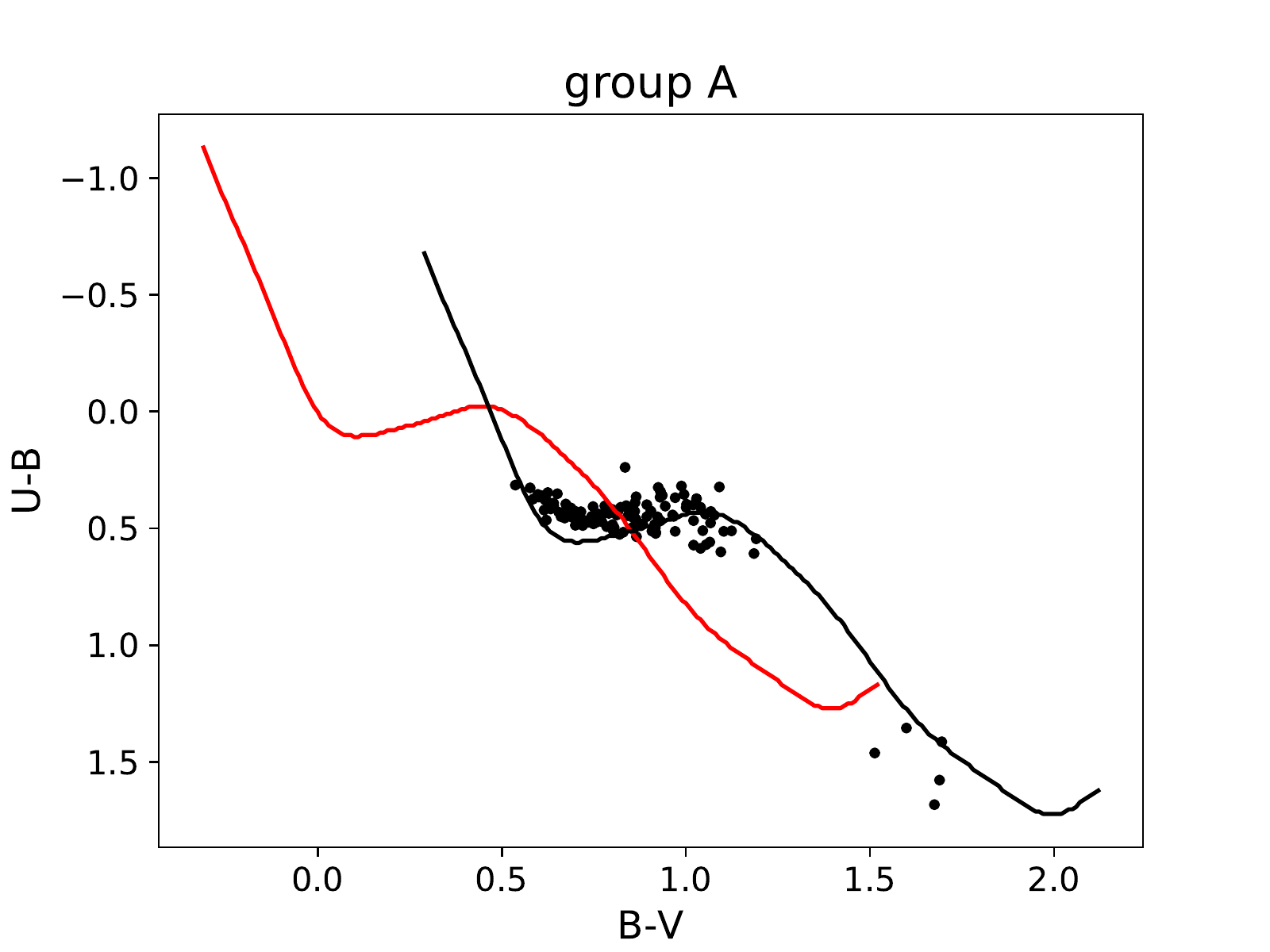}} \\
		\end{minipage}
		\hfill
		\begin{minipage}[h]{0.3\linewidth}
			\center{\includegraphics[width=1\linewidth]{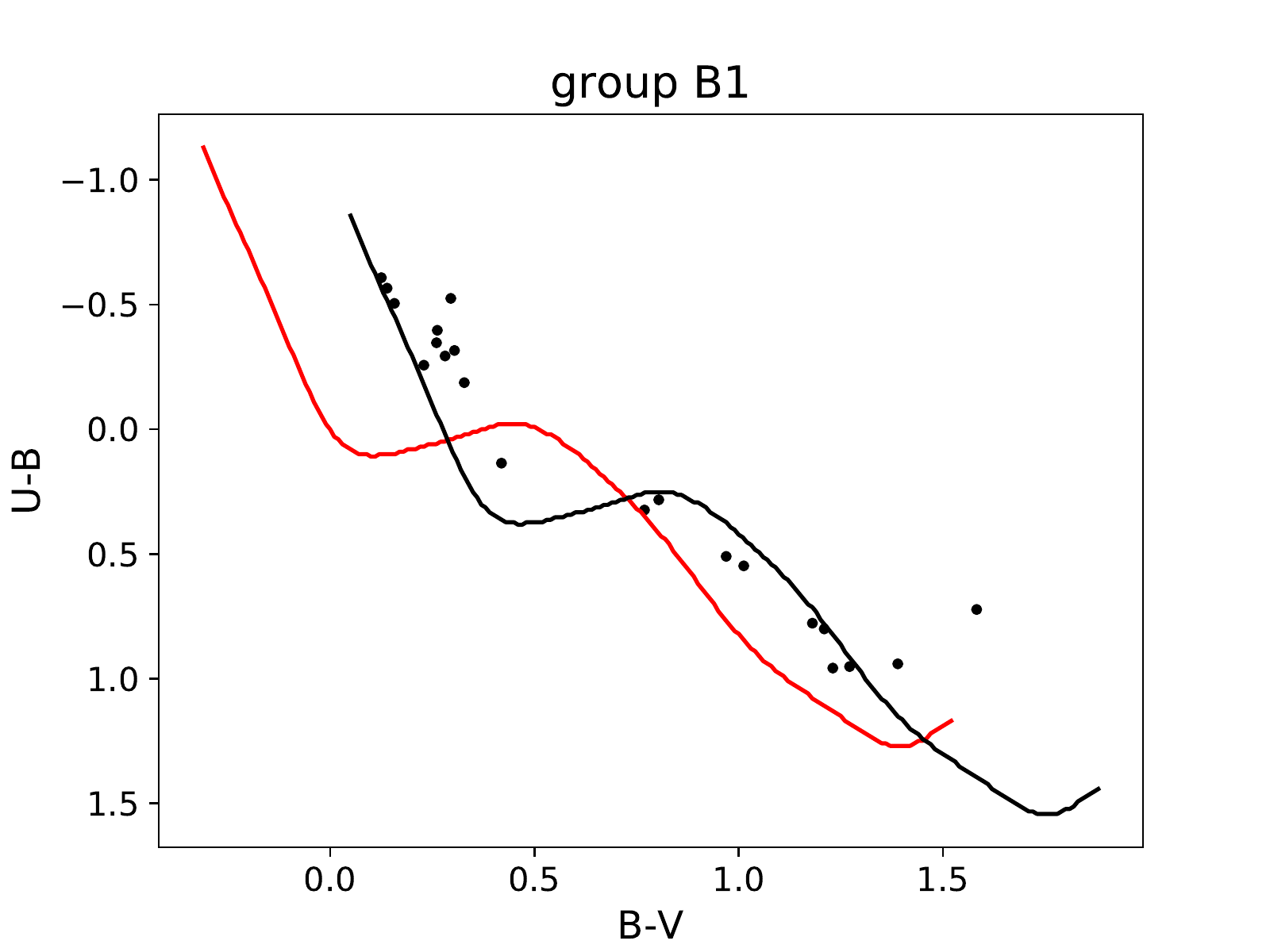}} \\
		\end{minipage}
		\hfill
		\begin{minipage}[h]{0.3\linewidth}
			\center{\includegraphics[width=1\linewidth]{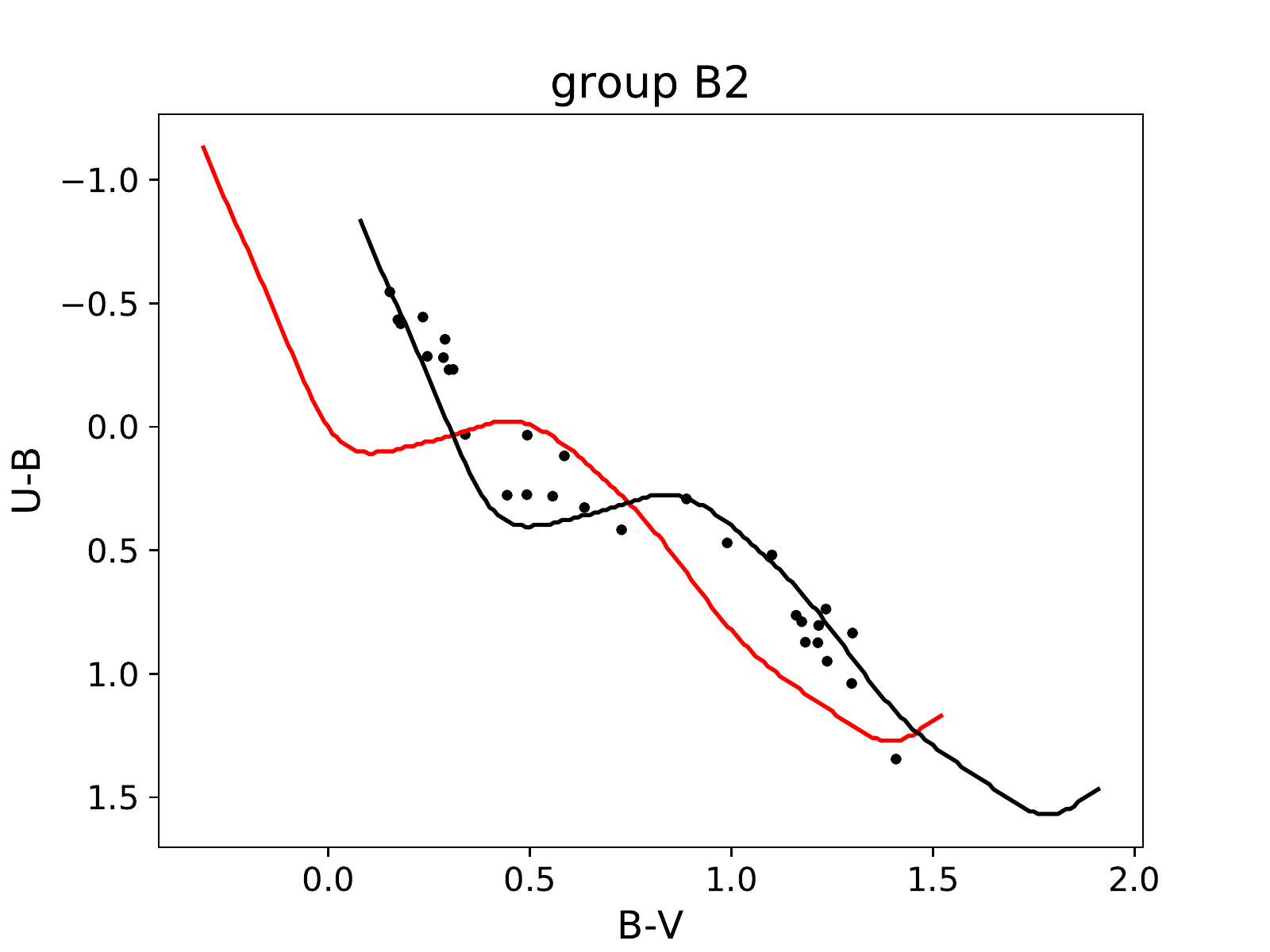}} \\
		\end{minipage}
		\vfill
		\begin{minipage}[h]{0.3\linewidth}
			\center{\includegraphics[width=1\linewidth]{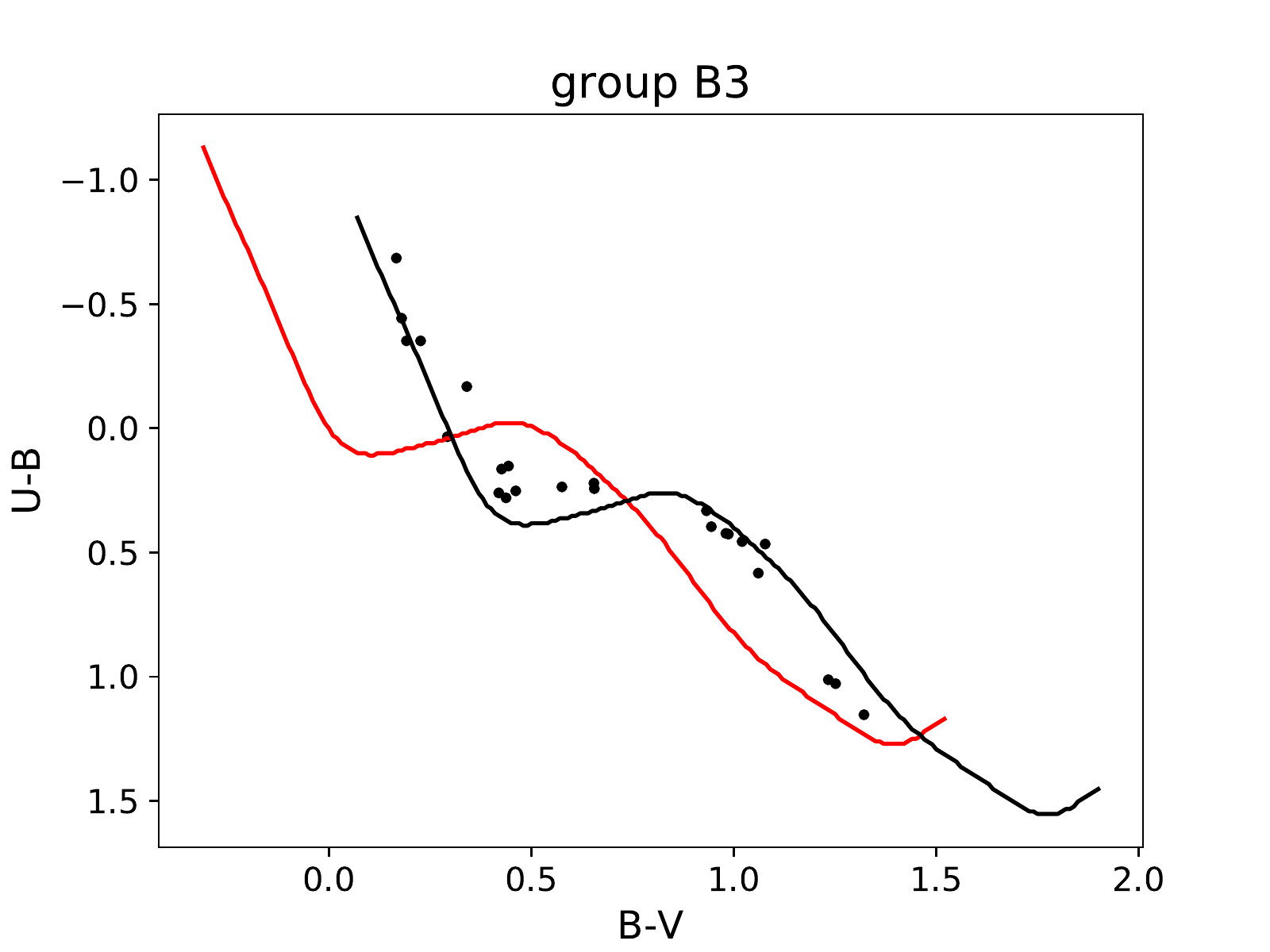}}\\
		\end{minipage}
		\hfill
		\begin{minipage}[h]{0.3\linewidth}
			\center{\includegraphics[width=1\linewidth]{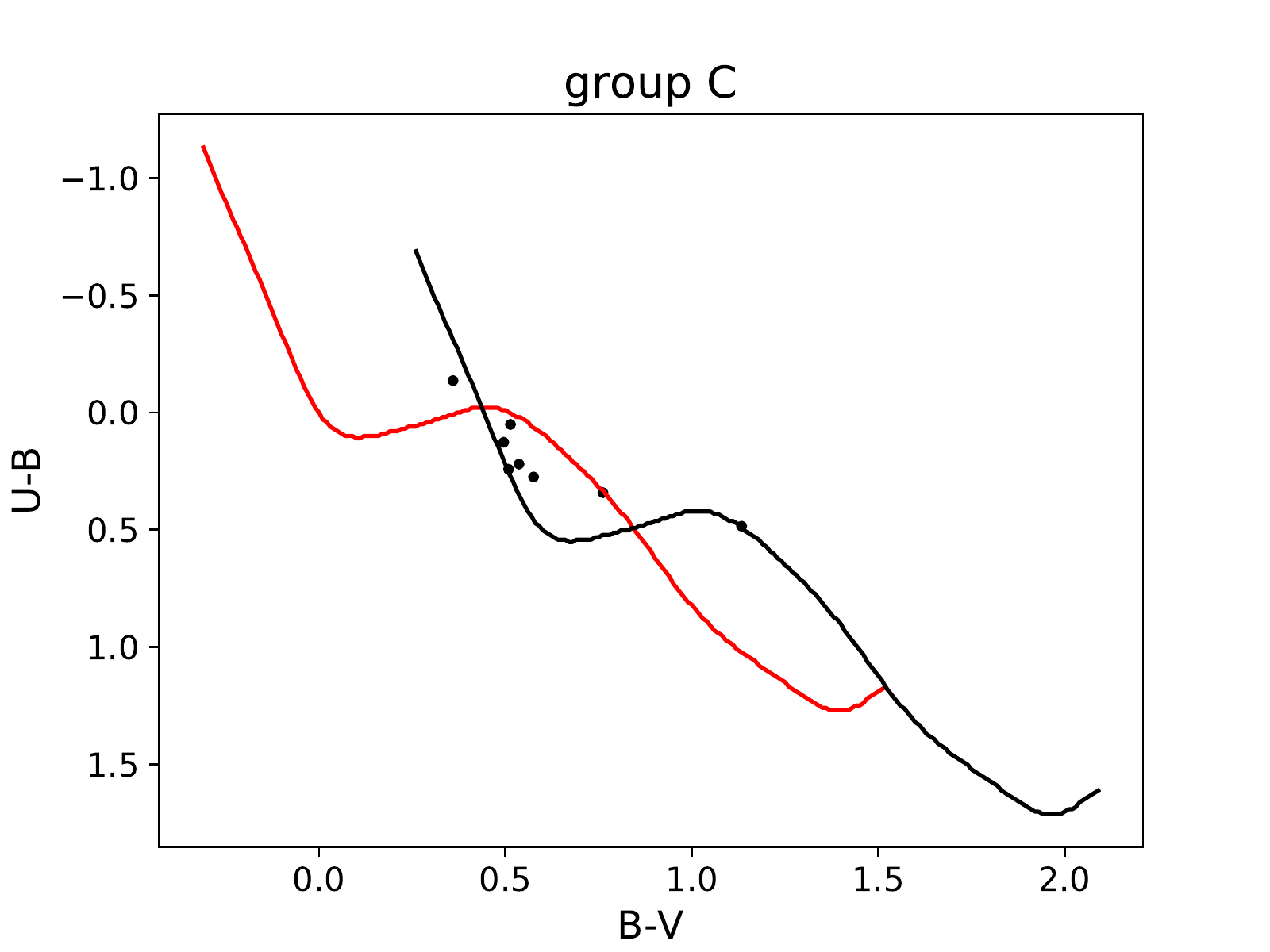}}\\
		\end{minipage}
		\hfill
		\begin{minipage}[h]{0.3\linewidth}
			\center{\includegraphics[width=1\linewidth]{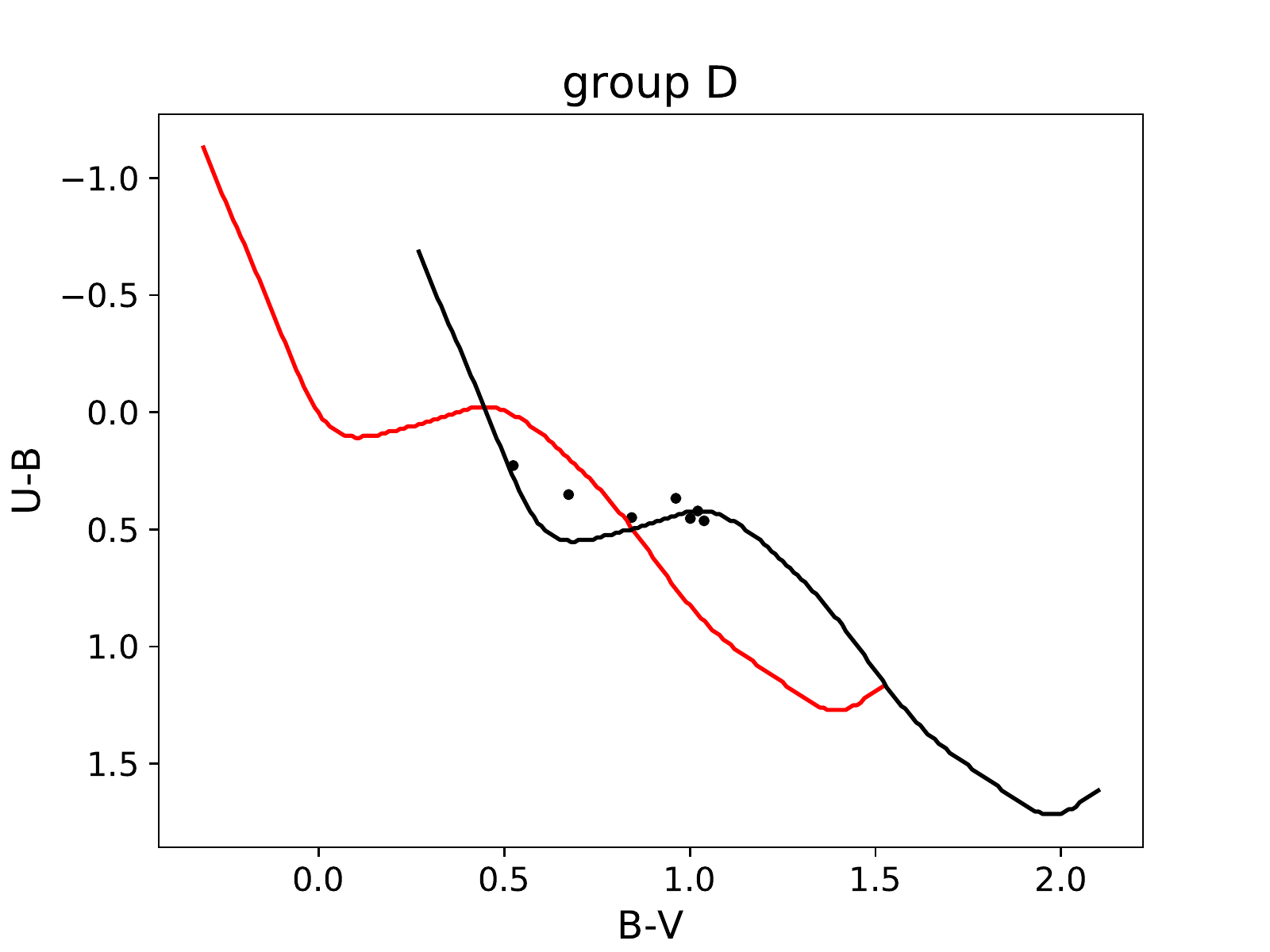}}\\
		\end{minipage}
		\vfill
		\begin{minipage}[h]{0.3\linewidth}
			\center{\includegraphics[width=1\linewidth]{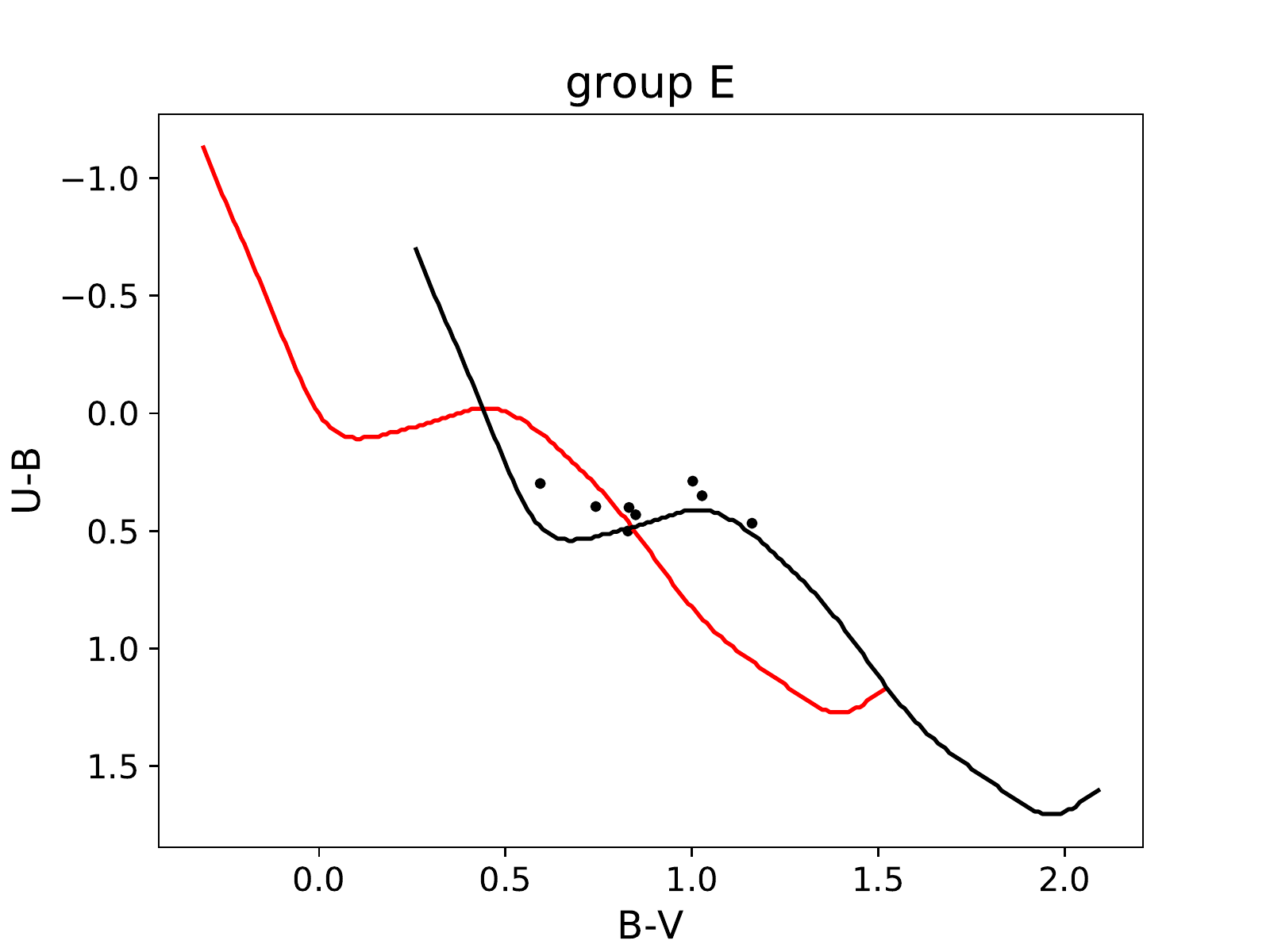}}\\
		\end{minipage}
		\hfill
		\begin{minipage}[h]{0.3\linewidth}
			\center{\includegraphics[width=1\linewidth]{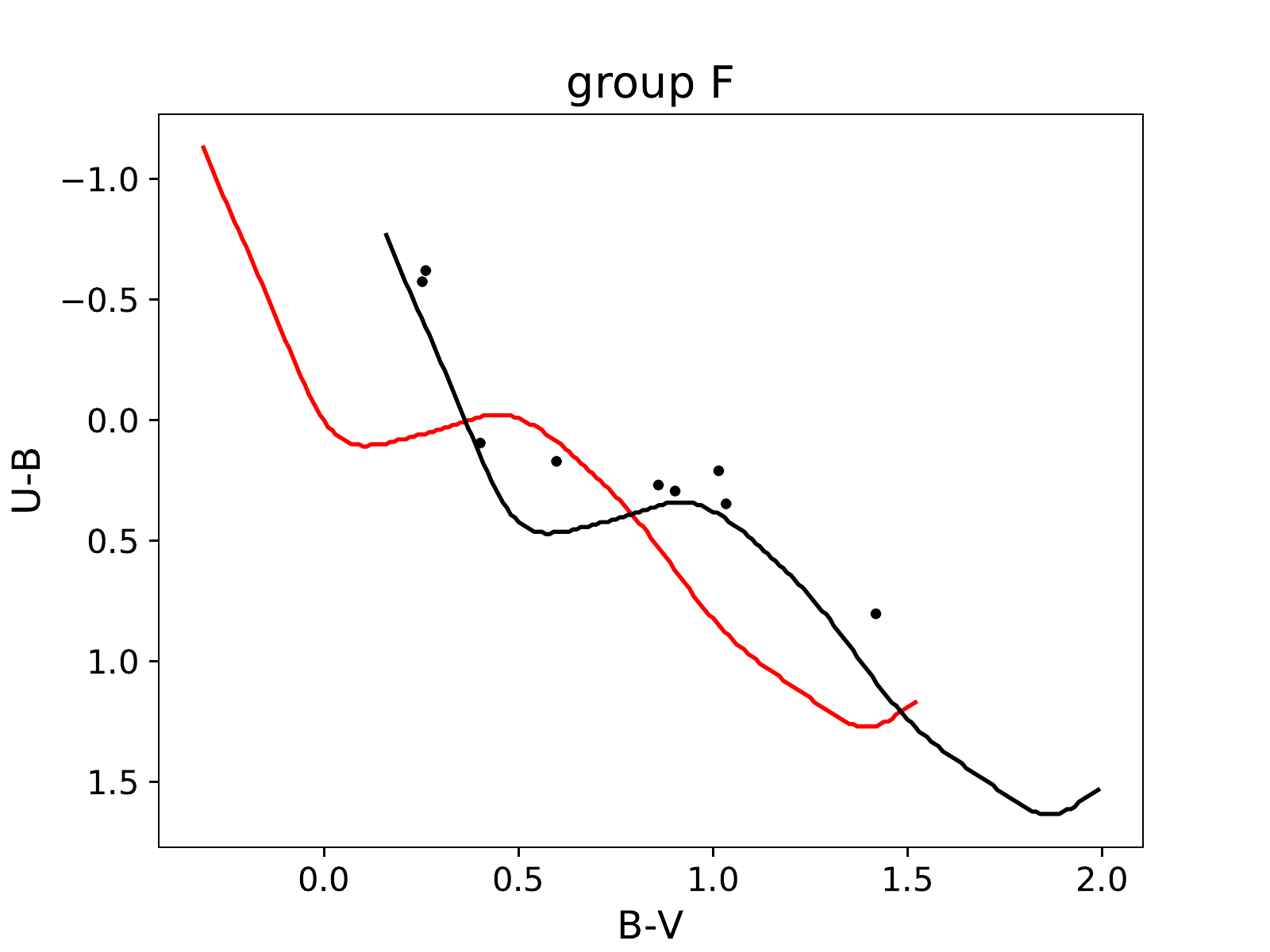}}\\
		\end{minipage}
		\hfill
		\begin{minipage}[h]{0.3\linewidth}
			\center{\includegraphics[width=1\linewidth]{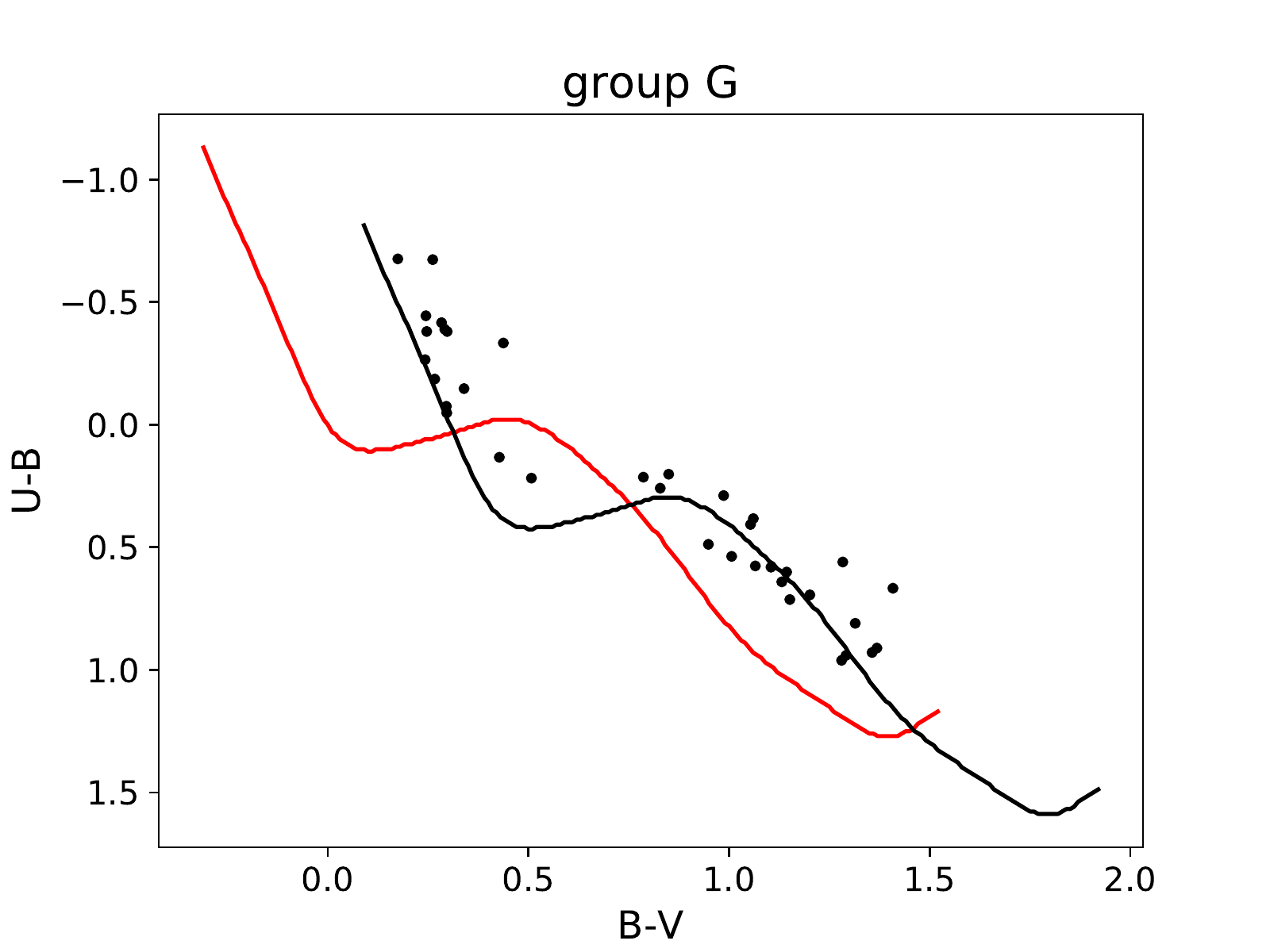}}\\
		\end{minipage}
		\caption{Color-color diagrams of groups. Red and black lines are intrinsic and shifted ZAMS.}
		\label{A1:ccd}
	\end{figure*}
	
	\section{Color-magnitude diagrams}
	
	\begin{figure*}
		\begin{minipage}[h]{0.3\linewidth}	
			\center{\includegraphics[width=1\linewidth]{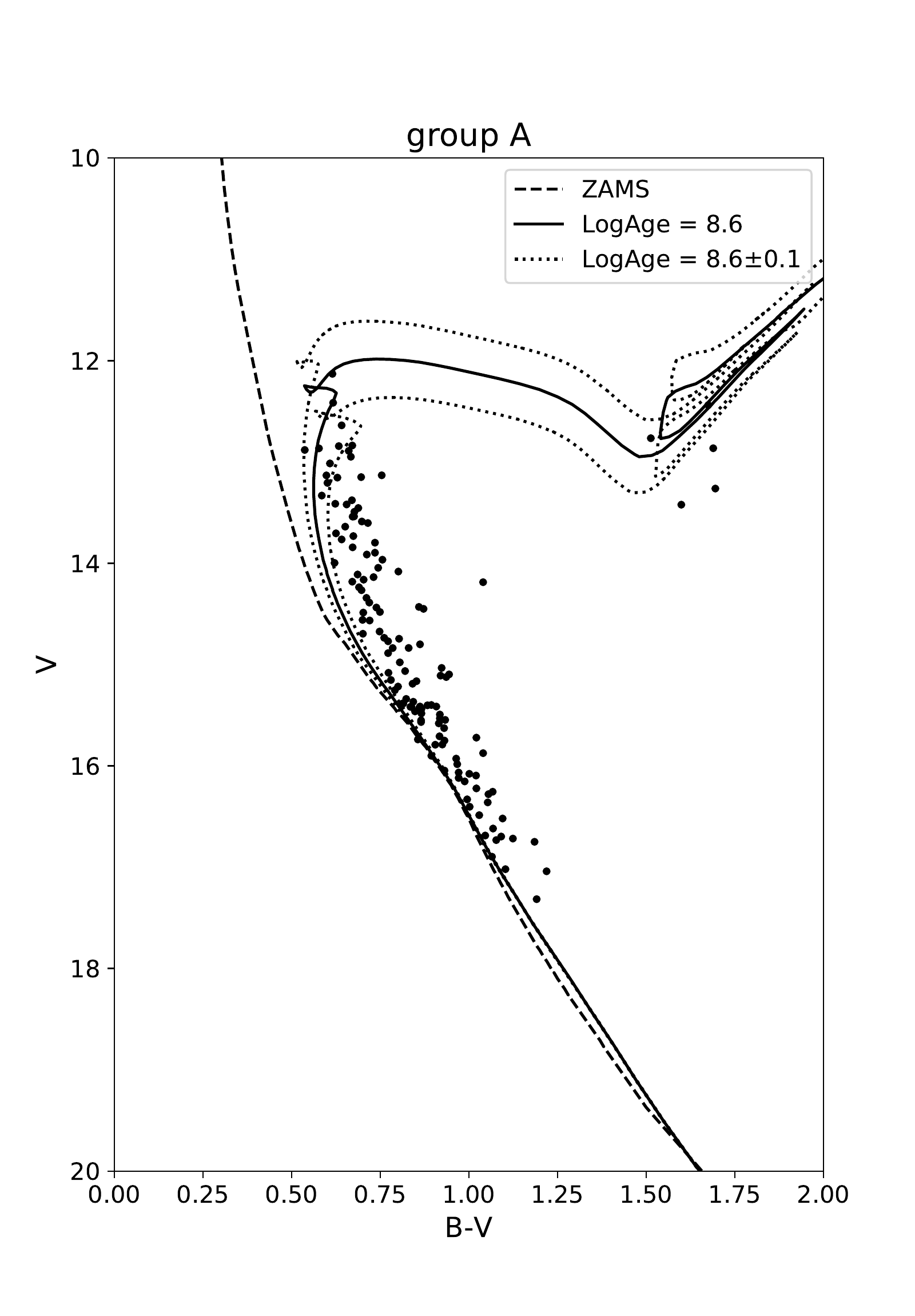}} \\
		\end{minipage}
		\hfill
		\begin{minipage}[h]{0.3\linewidth}
			\center{\includegraphics[width=1\linewidth]{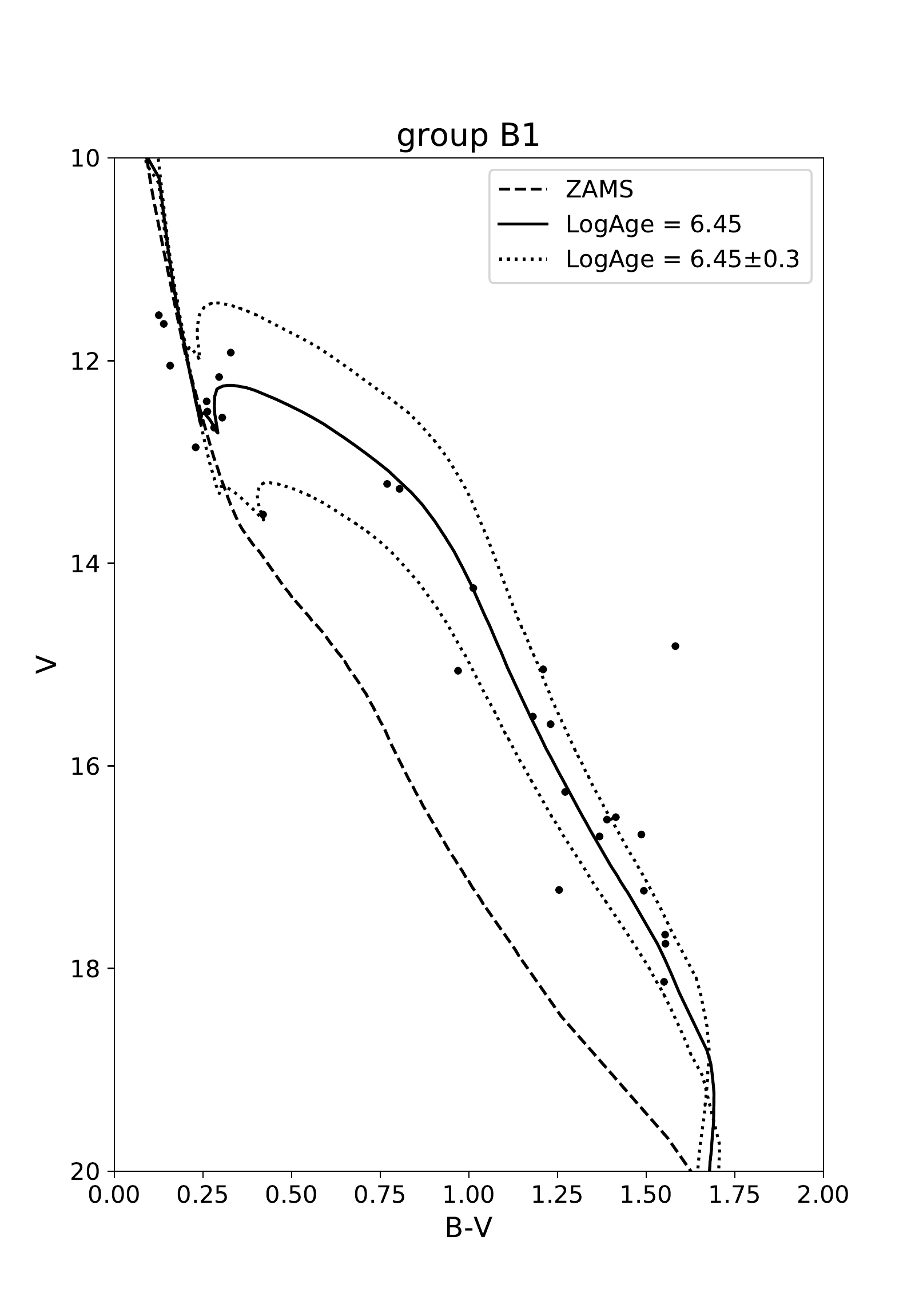}} \\
		\end{minipage}
		\hfill
		\begin{minipage}[h]{0.3\linewidth}
			\center{\includegraphics[width=1\linewidth]{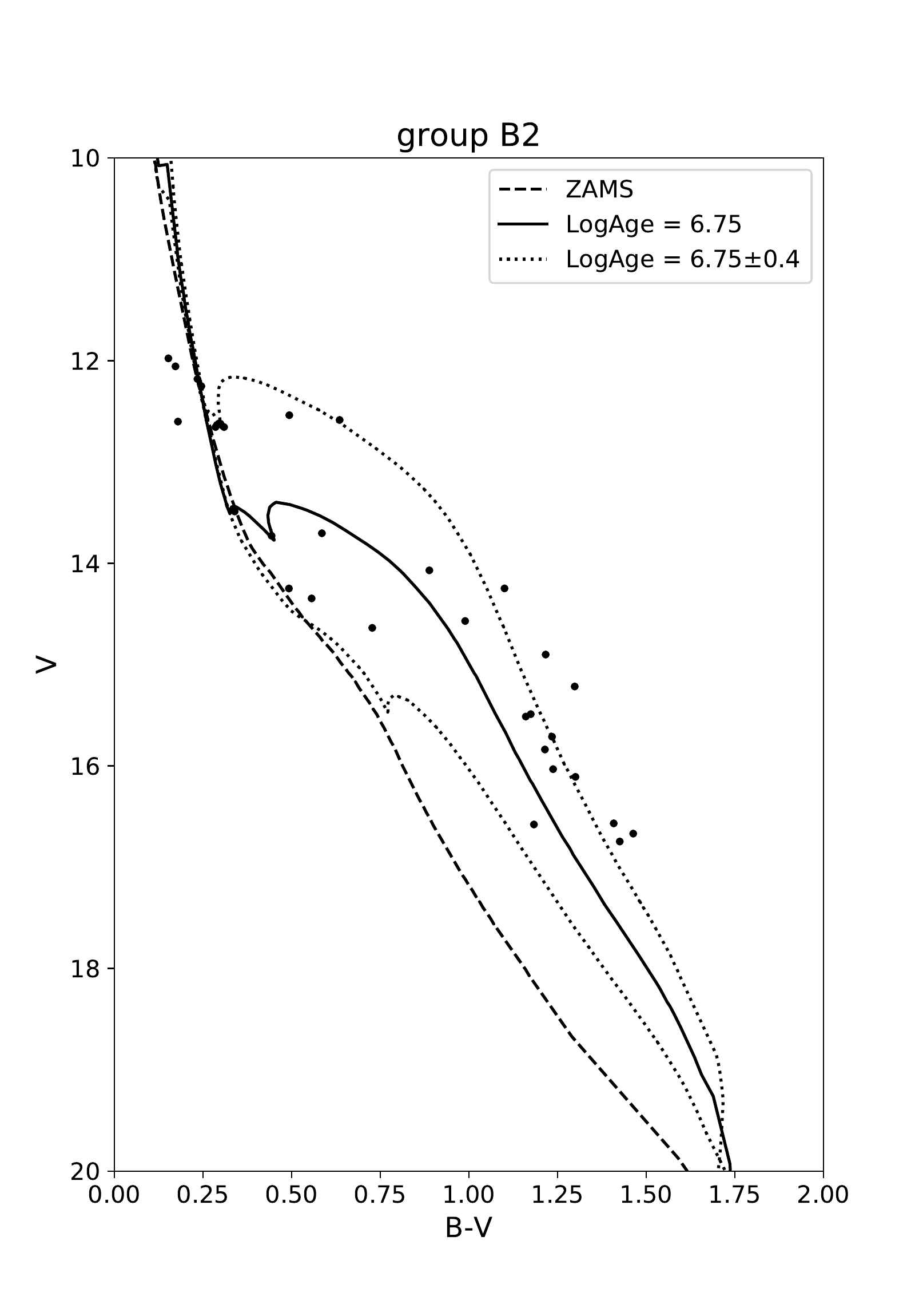}} \\
		\end{minipage}
		\vfill
		\begin{minipage}[h]{0.3\linewidth}
			\center{\includegraphics[width=1\linewidth]{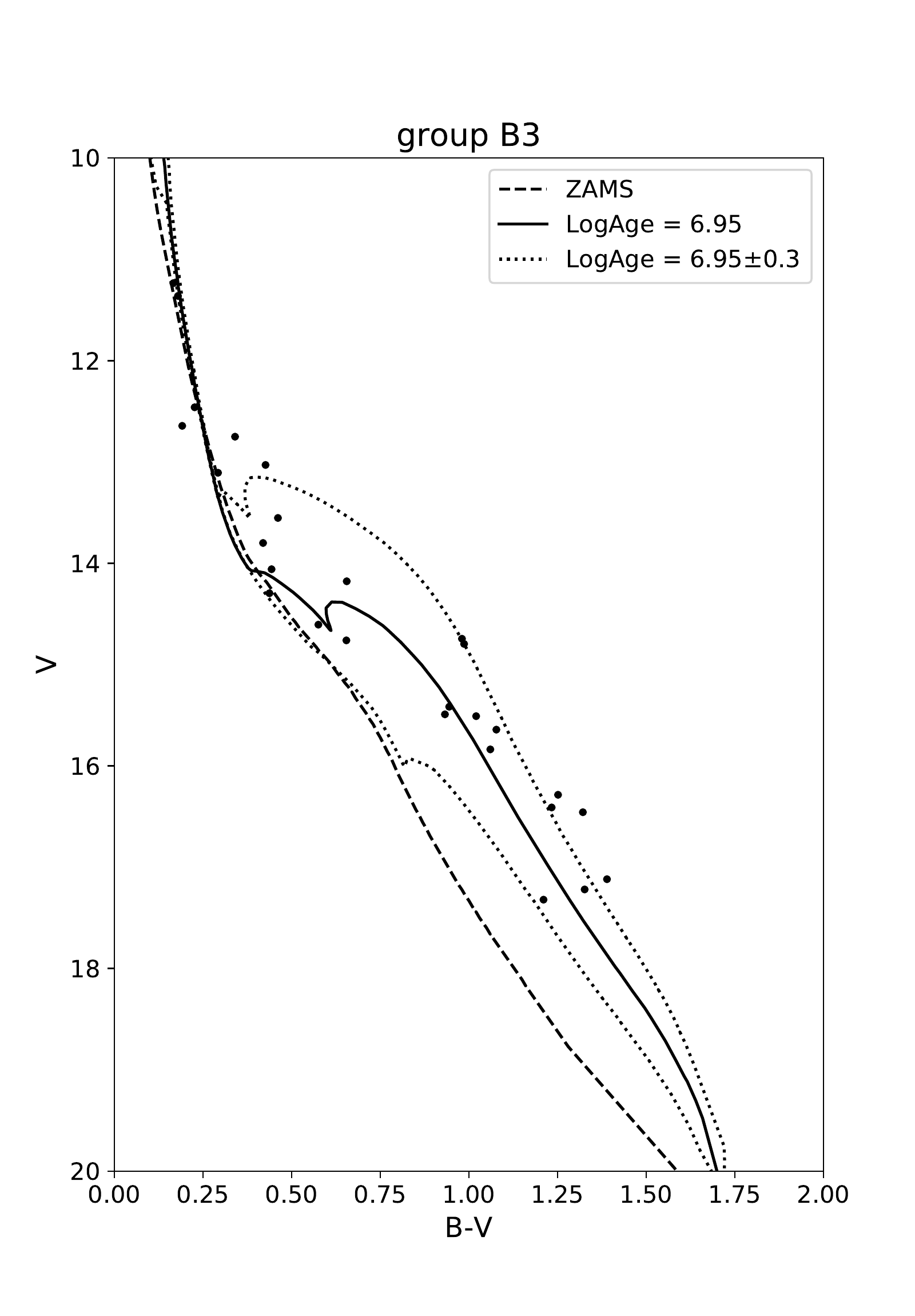}}\\
		\end{minipage}
		\hfill
		\begin{minipage}[h]{0.3\linewidth}
			\center{\includegraphics[width=1\linewidth]{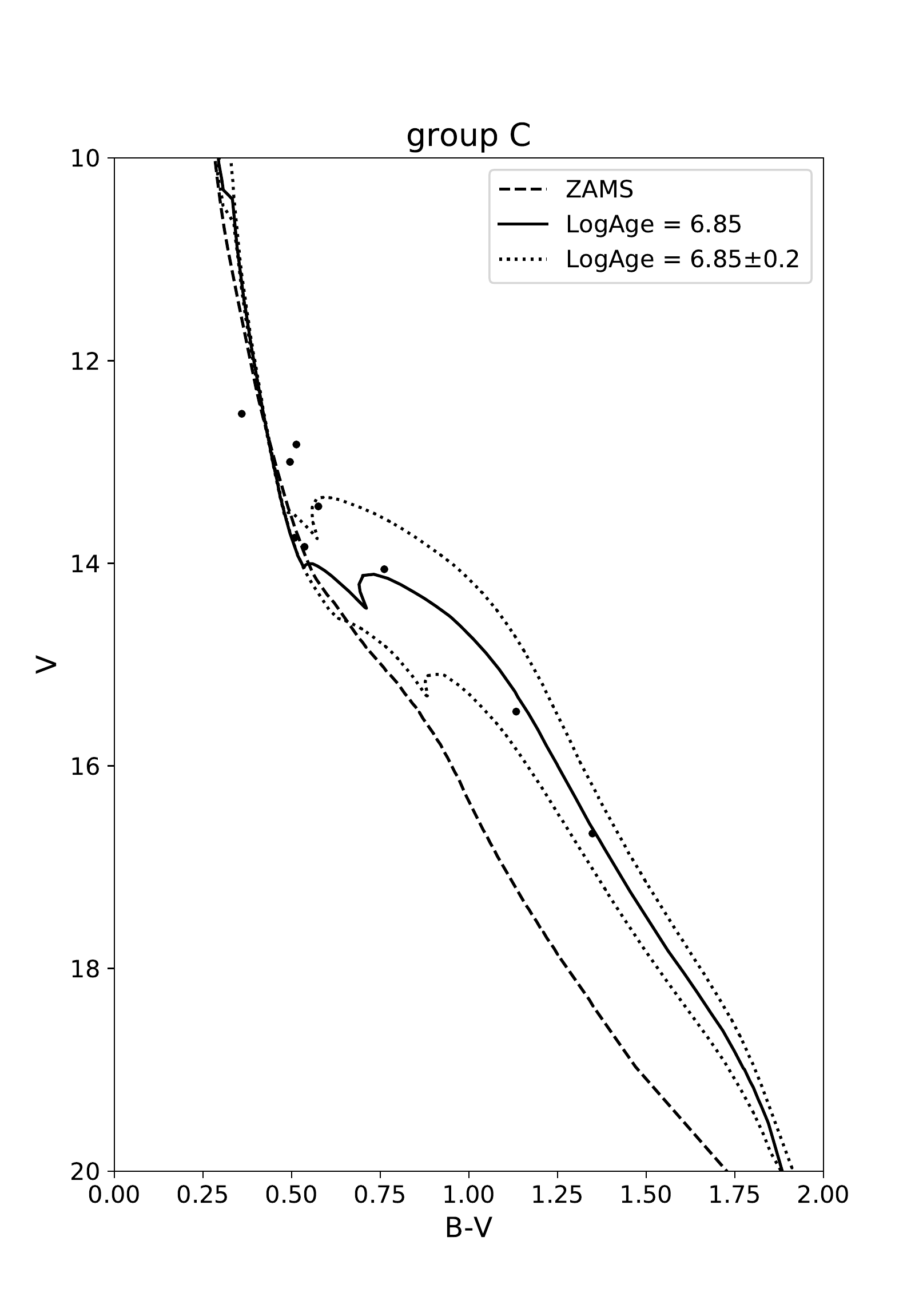}}\\
		\end{minipage}
		\hfill
		\begin{minipage}[h]{0.3\linewidth}
			\center{\includegraphics[width=1\linewidth]{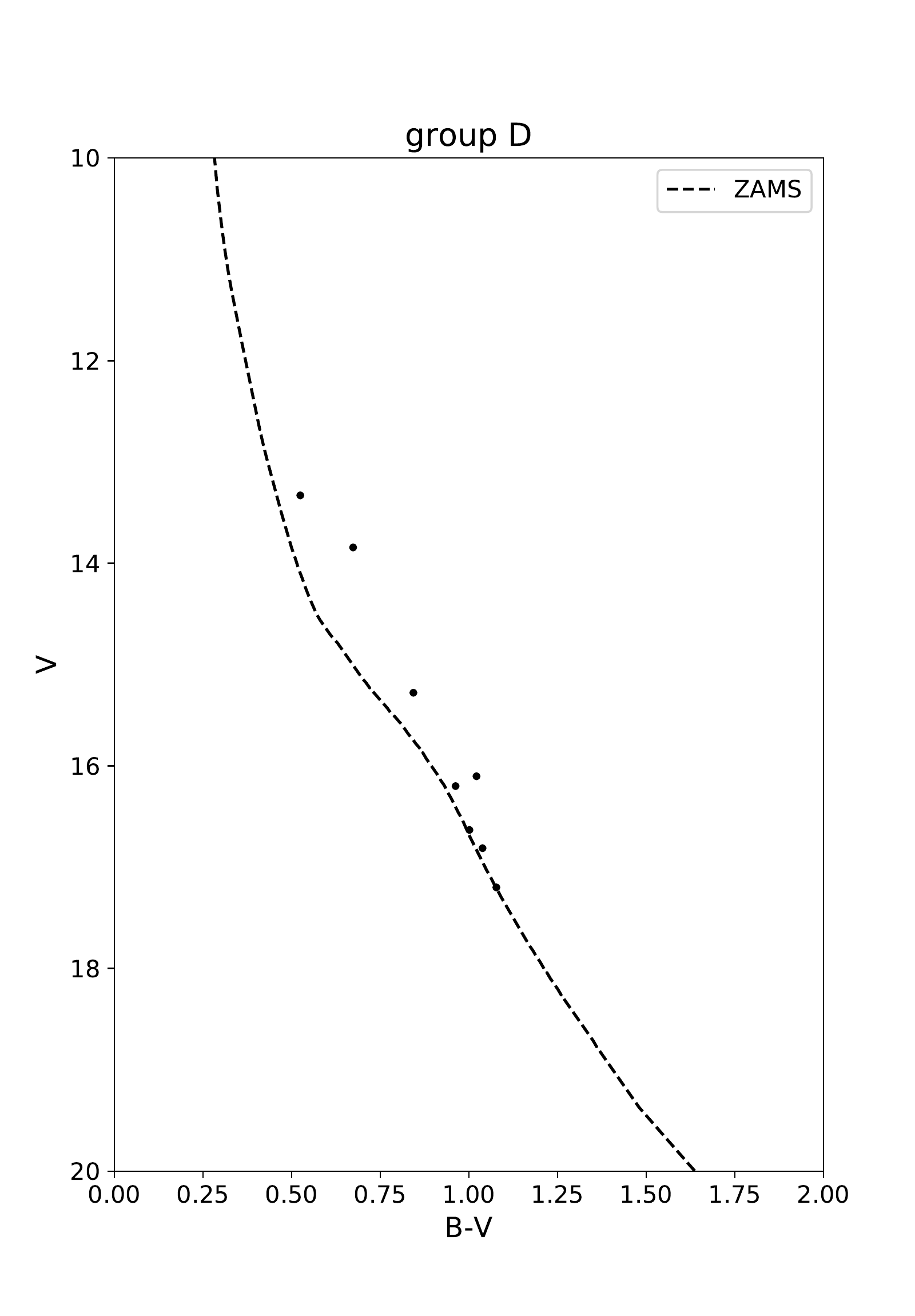}}\\
		\end{minipage}
		\vfill
		\begin{minipage}[h]{0.3\linewidth}
			\center{\includegraphics[width=1\linewidth]{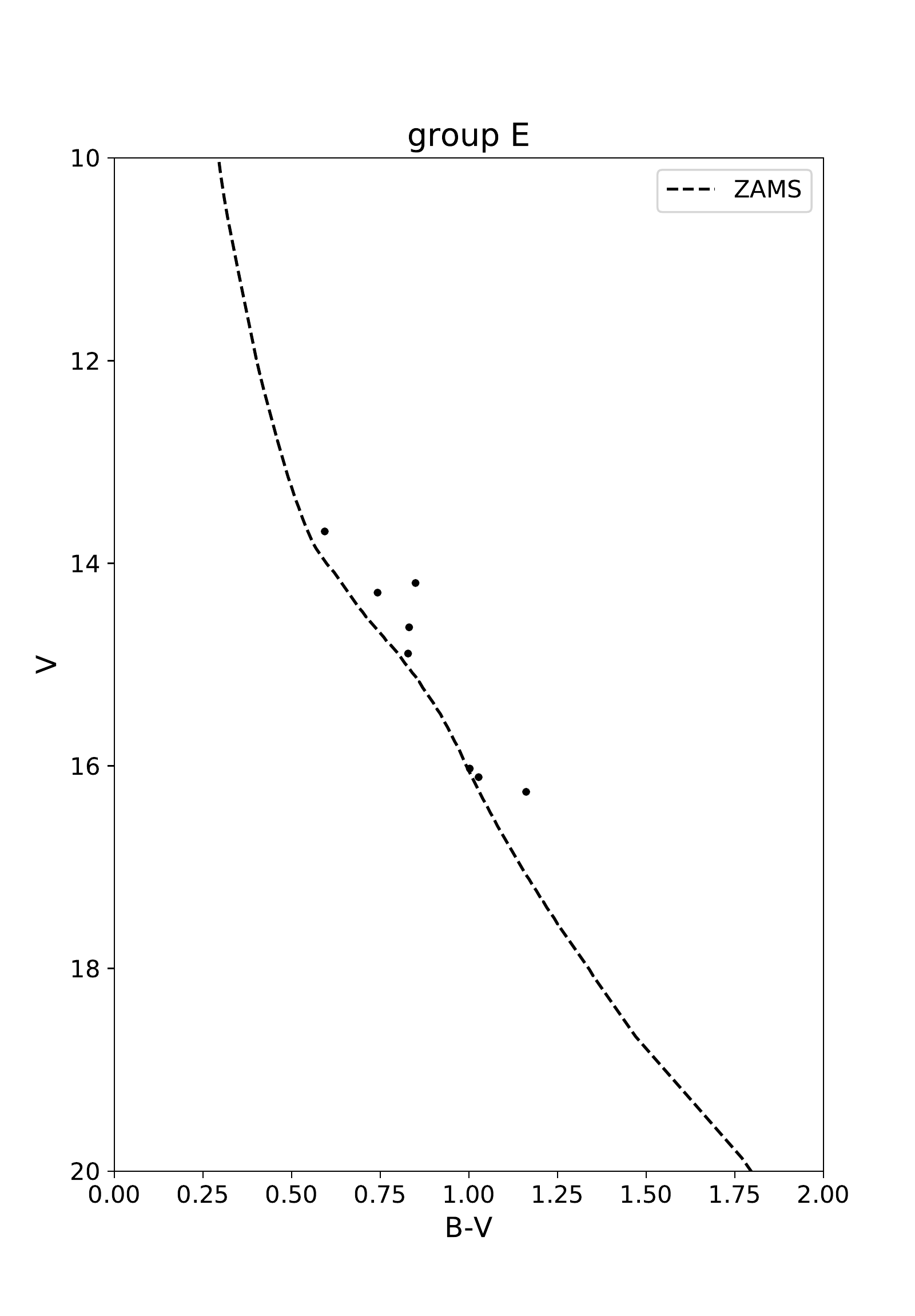}}\\
		\end{minipage}
		\hfill
		\begin{minipage}[h]{0.3\linewidth}
			\center{\includegraphics[width=1\linewidth]{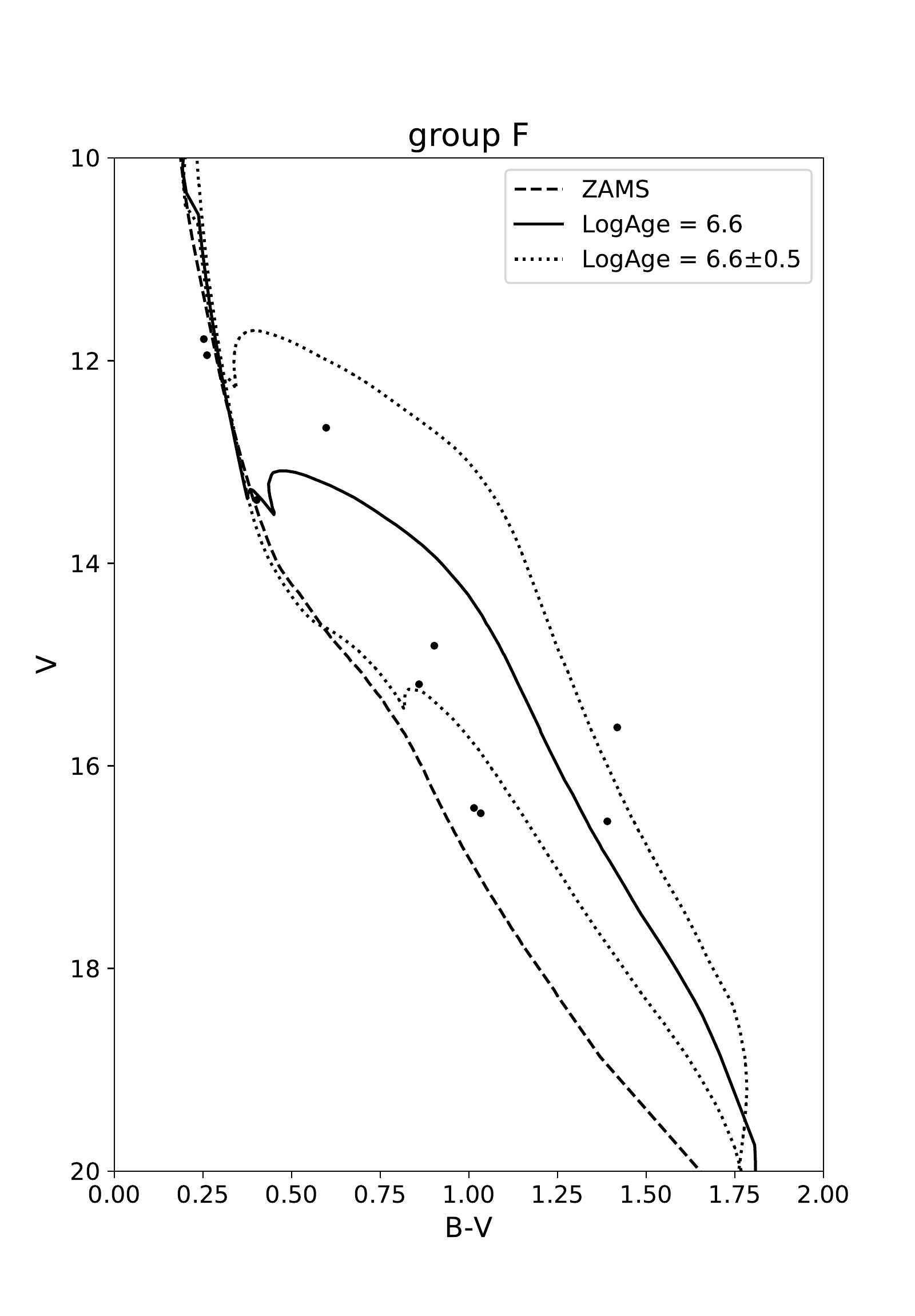}}\\
		\end{minipage}
		\hfill
		\begin{minipage}[h]{0.3\linewidth}
			\center{\includegraphics[width=1\linewidth]{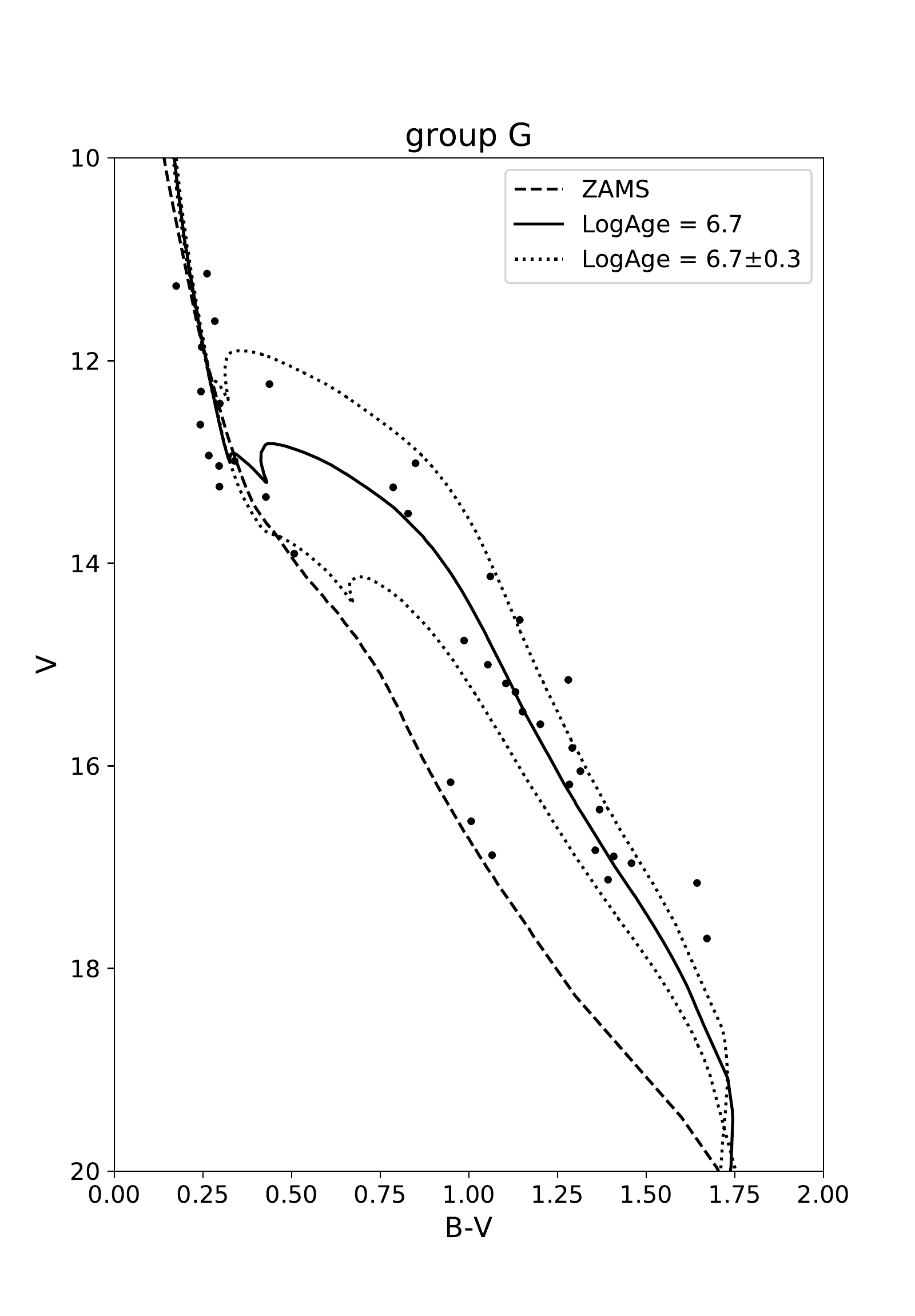}}\\
		\end{minipage}
		\caption{Color-magnitude diagrams of groups. Black dashed lines are shifted ZAMS. Solid lines are shifted isochrones. For groups D and E ages were estimated by brightest stars.}
		\label{A1:cmd}
	\end{figure*}
	

	\bsp	
	\label{lastpage}
\end{document}